\documentclass[prl,reprint,superscriptaddress,showkeys,twocolumn,amsmath,amssymbm]{revtex4-2}
\usepackage{graphicx}
\usepackage{comment}
\usepackage[colorlinks=true,allcolors=blue]{hyperref}
\usepackage{braket}
\usepackage[normalem]{ulem}
\usepackage{xr}
\usepackage{hyperref}
\usepackage{xr-hyper} 
\usepackage{extpfeil}
\usepackage[dvipsnames]{xcolor}

\DeclareGraphicsExtensions{.pdf,.ai,.jpg}
\begin{document}

\title{Consistent quantum treatments of nonconvex kinetic energies}

\author{C. Koliofoti, M. A. Javed and R.-P. Riwar}
\affiliation{Peter Gr\"unberg Institute, Theoretical Nanoelectronics, Forschungszentrum J\"ulich, D-52425 J\"ulich, Germany}

\begin{abstract}

The task of finding a consistent relationship between a quantum Hamiltonian and a classical Lagrangian is of utmost importance for basic, but ubiquitous techniques like canonical quantization and path integrals. Nonconvex kinetic energies (which appear, e.g., in nonlinear capacitors or classical time crystals) pose a fundamental problem: the Legendre transformation is ill-defined, and the more general Legendre-Fenchel transformation removes nonconvexity essentially by definition. Arguing that such anomalous theories follow from suitable low-energy approximations of well-defined, harmonic theories, we show that seemingly inconsistent Hamiltonian and Lagrangian descriptions can both be valid, depending on the coupling strength to a dissipative environment. There occurs a dissipative phase transition from a nonconvex Hamiltonian to a convex Lagrangian regime, involving exceptional points in imaginary time. Our approach thus resolves apparent inconsistencies and provides computationally efficient methods to treat anomalous, nonconvex kinetic energies.

\end{abstract}

\maketitle

\textit{Introduction. -- } Canonical quantization (CQ)~\cite{Schwabl2007} and the path integral (PI) formalism~\cite{Zee2010}, are indispensable tools for modern quantum theory. Both rely on a well-defined correspondence between a classical Lagrangian $L$ and a quantum Hamiltonian $H$, which is usually obtained from the Legendre transformation~\footnote{Strictly speaking, the connection between $H$ and $L$ for PI is related to a Fourier transformation, and not the Legendre transformation. The two are only equivalent for \textit{harmonic} kinetic energies.}. But there are various important cases where the Legendre transformation is ill-defined~\footnote{In line with the previous footnote, the conventional PI approach loses validity for \textit{any} anharmonic kinetic energy, an issue that is well-known in the relativistic community, see, e.g., Ref.~\cite{Padmanabhan_2018} and references therein.}~\cite{Padmanabhan_2018}. The velocity-momentum relationship is non-invertible if the mass tensor has zero eigenvalues, requiring alternative quantization procedures~\cite{Bergmann_1949,Faddeev_1988,Brown_2022,ParraRodriguez2024} which have recently been subject to controversy~\cite{Rymarz_2023,Egusquiza_2025comment,DiVincenzo_2025reply}. Shapere and Wilczek’s classical time crystal is based on the notion of nonconvex kinetic energies in $L$~\cite{Shapere_2012}, whose consistent quantization presents an unresolved challenge to this day~\cite{Zhao_2013,Dai_2020}, where in particular "branched quantization" methods~\cite{Henneaux_1987,Shapere_2012b,Choudhury_2019} have been critiziced~\cite{Chi_2014}.

This work focuses on nonconvex kinetic energies in the \textit{Hamiltonian}. Such Hamiltonians appear in the quantum circuit context as nonlinear, anomalous capacitors, either in the form of electrostatically coupled polarizer objects~\cite{Little_1964,Hamo_2016,Herrig_2023,Herrig_2025}, polarizing materials (ferroelectrics)~\cite{Landauer_1976,Catalan_2015,Ng_2017,Hoffmann_2018,Lukyanchuk_2019,Hoffmann_2020}, or quantum phase slip junctions~\cite{Giordano1988,Bezryadin2000,Lau_2001,Buchler2004,Mooij_2006,Astafiev_2012,deGraaf_2018,Shaikhaidarov_2022,Koliofoti_2023}. While the Hamiltonian is here known, the path to a corresponding Lagrangian is nonetheless of fundamental importance for a quantum description~\cite{Burkard_2004,Vool_2017,Riwar_2022} of circuit hardware with ever increasing complexity~\cite{Arute_2019,IBM_roadmap}, and in order to efficiently describe dissipation via PI~\cite{Altland_Simons_book}. 
Reconstructing a valid Lagrangian from a nonconvex Hamiltonian is problematic for the same reason as for classical time crystals~\cite{Chi_2014}: even though the Legendre-Fenchel transformation formally applies to nonconvex functions, it replaces them with their convex hull~\footnote{Specifically, the Lagrangian obtained from the Legendre-Fenchel-transformed Hamiltonian becomes in general non-smooth, and a second application does not recover the original nonconvex Hamiltonian, but a deformed, convexificated variant.}. 
While for quantum phase slip junctions, methods have been proposed to circumvent nonconvexity~\cite{Ulrich_2016,Koliofoti_2023}, a general understanding of nonconvex kinetic energies, and their efficient PI treatment, remain an unresolved fundamental problem.

We study models (which in special cases map to the quantum Rabi~\cite{Ashhab_2010,Ashhab_2013,Hwang_2015} and Dicke models~\cite{Hwang_2016,Peng_2019}) where anomalous kinetic energies arise from suitable low-energy approximations of regular, harmonic theories. This allows us to derive seemingly inconsistent $H$ and $L$, where (in qualitative, but generally not quantitative, agreement with Ref. \cite{Chi_2014}) the Lagrangian completely misses the phase transition from convex to nonconvex (which, for the Rabi and Dicke models, is known as the superradiant transition). Surprisingly, when including dissipation, we find that both results can be correct  -- in complementary regimes, marking an additional \textit{dissipative} phase transition. We show that nonconvexity in $H$ coincides with \textit{exceptional points} in imaginary time PI. At low dissipation, imaginary time paths reach the exceptional points, such that the nonconvex $H$ is correct, whereas the low-energy approximation of $L$ fails due to adiabaticity breaking. Conversely, sufficiently strong dissipation smoothens out the imaginary time paths, restoring adiabaticity, validating the convex solution due to $L$. Overall, we demonstrate that inconsistent Hamiltonian and Lagrangian treatments both have their validity depending on dissipation strength, and provide a computationally efficient PI treatment of anomalous kinetic energies. 

Our work connects in unexpected ways to other ongoing research thrusts. Exceptional points are currently being explored as a means to describe topological phase transitions in open, driven quantum systems \cite{Ding_2018,Riwar_2019b,Fahri_2021,Liao_2021,Hlushchenko_2021,Shiqiang_2021,Ding_2022,Javed_2023,Kunst:2018aa,Kawabata:2019aa,Heiss_2004,Heiss_2012,Wu_2019,Mandal_2021,Bergholtz_2021,Avila_2019,San-Jose_2016,Khandelwal_2021,Khandelwal_2024}. While the transition from convex to nonconvex energies occurs without gap closing in the \textit{real-time} energy spectrum, exceptional points emerge here for \textit{imaginary} times. Dissipative phase transitions in the Caldeira-Legget model~\cite{Schmid_1983,Bulgadaev_1984} and the spin boson model~\cite{Leggett_1987} have been proposed long ago, but especially for the former, experimental verification~\cite{Murani_2020,Hakonen_2021,Murani_2021,kuzmin2023observation,houzet2024microwave,burshtein2024inelastic} and some theoretical aspects~\cite{Masuki_2022,Sepulcre_2023,Masuki_2023,giacomelli2024emergent,kashuba2024phasetransitions} remain controversial to this day. The dissipative phase transition proposed here is similar in spirit, but, crucially, relies on a conceptually simpler mechanism, in that the energy barrier between two minimas is not renormalized by high frequency paths, but effectively disappears due to the environment-induced path adiabaticity. Finally, we expect our approach to provide a well-defined conceptional access to consistent a quantum treatment of classical time crystals.

\textit{Minimal toy model. -- }Before presenting our arguments in their most general form, we begin for simplicity with a concrete minimal toy model. While the here developed concepts hold independent of the specific physical platform, we stick throughout this work to the language of quantum circuits, where the pair of canonically conjugate operators are the Cooper pair charge $N$ and superconducing phase $\phi$. The velocity $\dot{\phi}$ corresponds to the voltage across the capacitor, as per second Josephson relation. 

\begin{figure}
    \centering
    \includegraphics[width=0.9\linewidth]{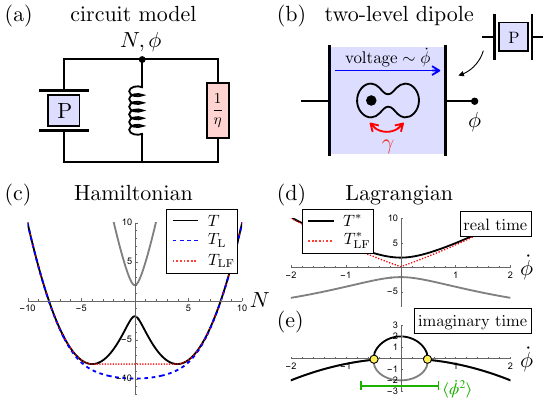}
    \caption{Main findings of this work, illustrated by simple toy model. (a) Nonlinear capacitor (with polarizer P) is shunted with an inductor and resistor. (b) Quantum dipole with two charge states (coupled with rate $\gamma$) interacting with the voltage across the capacitor. The Hamiltonian and Lagrangian descriptions of the device in (b) can be understood in terms of eigenvalues (black and grey lines) of $\widehat{T}(N)$ (c) and $\widehat{T}^*(\dot{\phi})$ (d), respectively, $\widehat{T}^*(i\dot{\phi})$ for imaginary time (e). The black lines mark the low-energy approximation. Blue dashed and red dotted lines correspond to Legendre-, respectively Legendre-Fenchel-transformed kinetic energies (see main text). (e) Exceptional points (marked with yellow dots) are relevant if PI voltage noise (green bar) is sufficiently large. In (e) only real part of the spectrum is plotted. In (c,d,e), the y-axis is energy in units of $1/c$, and $\lambda=4$, $c\gamma=2$.  }
    \label{fig1}
\end{figure}

Consider a parallel shunt of capacitor with an inductor. The capacitor (representing the kinetic energy) receives an effective nonlinearity via coupling to an intrinsic pseudo-spin degree of freedom (similar to Ref.~\cite{Placke_2018}). We will describe this system with both a Hamiltonian $H$ and a corresponding 'half-classical' Lagrangian $L$, where we only Legendre-transform with respect to $\dot{\phi}$~\footnote{The meaning and usefulness of such an incomplete Legendre transformation, especially in the context of the  PI treatment, will become clear throughout this work.}, 
\begin{equation}\label{eq_Lag_and_Hamil}
    H=\widehat{T}(N)+V(\phi) \qquad L=\widehat{T}^*(\dot{\phi})-V(\phi)
\end{equation}
with the kinetic energies $\widehat{T}(N)=\left(N+\lambda\sigma_z\right)^2/(2c)+\gamma\sigma_x$ and  $\widehat{T}^*(\dot{\phi)}=c\dot{\phi}^2/2-\lambda \dot{\phi} \sigma_z-\gamma \sigma_x$, where $\sigma_{x,z}$ are ordinary Pauli matrices. Note that we apply the standard Legendre transformation~\footnote{By standard Legendre transformation, we mean $N=\partial_{\dot{\phi}}L$ and $H=N\dot{\phi}-L$ when going from $L$ to $H$, and vice versa from $H$ to $L$.} to matrix-valued kinetic energies (due to $\sigma_{x,z}$). Since $L$ is quadratic (and linear) in $N$, respectively $\dot{\phi}$, this is a well-defined procedure. The shunt inductor is expressed with the potential energy term $V(\phi)$ (which is, e.g., $\sim \phi^2$ for a linear inductor, or $\sim\cos(\phi)$ for a Josephson junction~\footnote{For simplicity, this work will mostly focus on linear inductors, where the problem of nonconvex $T(N)$ could in principle be avoided by performing a Legendre transformation on $V(\phi)$ instead of $T(N)$, similar in spirit to Ref.~\cite{Ulrich_2016}. But our goal is to provide a theory that works independent of whether the inductive shunt is linear or nonlinear.}).
The bare capacitor has a capacitance $c$ (in units of inverse energy). Physically, the pseudo-spin can be regarded as a quantum dipole moment (e.g., a single electron trapped in a double quantum dot), where the voltage across the capacitor $\sim \dot{\phi}$ leads to a detuning $\sim \lambda\dot{\phi}$ and $\gamma$ expresses a tunneling process between the two dipole configurations, see Fig.~\ref{fig1}(b).

The pseudo-spin provides an energy gap $\sim\gamma$ in both $L$ and $H$, which allows us eliminate it via a low-energy approximation in Born-Oppenheimer style. This yields a scalar low-energy theory (where we drop the hat notation, $\widehat{T}\rightarrow T$). For $H$, one chooses the lowest eigenvalues with respect to spin [black line in Fig.~\ref{fig1}(c)],
\begin{equation}\label{eq_TH_low}
    \widehat{T}(N)\approx T(N)=\frac{1}{2c}N^2-\sqrt{\lambda^2N^2/c^2+\gamma^2}\ .
\end{equation}
For the Lagrangian on the other hand, we have to select the highest eigenvalue~\footnote{Note that the regular $\sim\cos(\phi)$ energy contribution from a standard Josephson tunneling junction follows from the exact same low-energy approximation of the full BCS description of the junction, within the Lagrangian. The reason for selecting the highest, and not the lowest eigenvalue becomes clear when considering the system in a thermal Gibbs state (for which we will use imaginary times), see further below.} [black line in Fig.~\ref{fig1}(d)],
\begin{equation}\label{eq_TL_low}
    \widehat{T}^*(\dot{\phi})\approx T^*(\dot{\phi})= \frac{c}{2}\dot{\phi}^2+\sqrt{\lambda^2 \dot{\phi}^2+\gamma^2}\ .
\end{equation}
The scalar kinetic energy $T$ in Eq.~\eqref{eq_TH_low} undergoes a phase transition at the critical point $\gamma_0=\lambda^2/c$. For $\gamma<\gamma_0$ ($\gamma>\gamma_0$), the kinetic energy is \textit{nonconvex} (convex), qualitatively resembling the known behaviour of ferroelectric materials~\cite{Landauer_1976,Catalan_2015,Ng_2017,Hoffmann_2018,Lukyanchuk_2019,Hoffmann_2020}. This transition can be probed, e.g., by the heat capacity, $C_v=\beta^2\partial_\beta^2\ln(Z)$ with $\beta$ the inverse temperature ($k_B=1$), and the partition function $Z=\text{tr}[e^{-\beta H}]$. For a linear inductive shunt, $V(\phi)=\phi^2/(2l)$ (where Eq.~\eqref{eq_Lag_and_Hamil} maps onto the Rabi model~\cite{Ashhab_2010,Ashhab_2013,Hwang_2015}, and above transition is known as the superradiant transition), the heat capacity can be computed analytically in the limit $\gamma \gg 1/\beta \gg 1/l$~\cite{Supplemental_Material}. In both the convex and nonconvex phases we find $C_v=1$, whereas $C_v=3/4$ at the critical point (due to $T\sim N^4$).

In contrast to $T$, the low-energy Lagrangian kinetic energy $T^*$, Eq.~\eqref{eq_TL_low}, always remains convex. In fact, for low voltages $\dot{\phi}\approx 0$, we can always approximate it as harmonic, $T^*\approx c_\text{eff}\dot{\phi}^2$. A trace of the phase transition remains in the form of a renormalization of the effective capacitance $c_\text{eff}= c+\lambda^2/\gamma$. This renormalization crosses over from weak before the transition ($\gamma \gg \gamma_0$),  $c_\text{eff}\approx c$, and strong after the transition ($\gamma \ll \gamma_0$), $c_\text{eff}\approx \lambda^2/\gamma \gg c$.

Because $T^*$ is always convex, we can in principle apply a standard Legendre transformation directly for the Lagrangian low-energy theory. The resulting Hamiltonian kinetic energy, denoted as $T_\text{L}(N)$, \textit{cannot} possibly be consistent with the direct low-energy $T(N)$ in Eq.~\eqref{eq_TH_low}, as it is convex by definition, thus missing the superradiant phase transition. We can get a third (yet again inconsistent) version of the low-energy kinetic energy in $H$, by applying the Legendre-Fenchel transformation to the original, nonconvex $T(N)$, arriving at an intermediary Lagrangian with $T^*_\text{LF}(\dot{\phi})$, and applying it a second time to get $T_\text{LF}(N)$. $T_\text{L}$ and $T_\text{LF}$ are qualitatively similar (both convex), but only the former remains smooth, see Fig.~\ref{fig1}(c) and (d) (in alignment with Ref. \cite{Chi_2014}).

Cross-checking with the full model, Eq.~\eqref{eq_general_model}, it is easy to verify that $T$ is correct (at least under the assumptions so far stated), whereas $T_\text{L}$ and $T_\text{LF}$ must be rejected. While $T_\text{LF}$ can (at least in the present context) be discarded as a mere mathematical exercise, the inconsistency between $T$ and $T_\text{L}$ is more surprising as both follow from a low-energy approximation due to an energy gap. In particular, had we given the convex low-energy $L$ in Eq.~\eqref{eq_TL_low} as starting point, we would not have found the correct, nonconvex low-energy $H$. To our knowledge, no work in the existing literature warns about the possibility that CQ (applying the standard Legendre transformation) for an anharmonic, but strictly convex $T^*$ in $L$ could provide a wrong Hamiltonian. 

\textit{Imaginary time path integrals. --} To understand the failure of $T_\text{L}$, we turn to the PI approach. Since we consider finite temperatures, we turn to imaginary times
, expressing the partition function for the full system as~\cite{Supplemental_Material}
\begin{equation}\label{eq_Z_PI}
    Z=\oint \mathcal{D}\phi \text{tr}_{\sigma}\left[\mathcal{T}e^{\int_0^\beta d\tau \left[\widehat{T}^*(i\dot{\phi}_\tau)-V(\phi_\tau)\right] }\right]\ ,
\end{equation}
where the notation $\oint\mathcal{D}\phi$ represents the ordinary functional integral over all possible paths $\phi_\tau$ with periodic boundary conditions $\phi_\beta=\phi_0$~\cite{Altland_Simons_book}. We only transformed the pair of $N,\phi$ into classical variables $\dot{\phi},\phi$ (while leaving the pseudo-spin quantum). Therefore, there is a separate trace over spin ($\text{tr}_\sigma[\ldots]$), and the imaginary 'time evolution' requires time ordering ($\mathcal{T}$).  

Due to $t\rightarrow -i\tau$, the voltage gets a prefactor $i$, and the eigenvalues of $\widehat{T}^*$ are $-c\dot{\phi}_{\tau}^2/c\pm\sqrt{\gamma^2-\lambda^2\dot{\phi}_\tau^2}$. Consequently, when $\dot{\phi}_\tau$ passes the threshold value $\gamma/\lambda$, the system exhibits \textit{exceptional points}. In contrast, both the Hamiltonian and \textit{real-time} Lagrangian always have a well-defined direct energy gap with respect to the pseudo-spin. This explains the failure of the low-energy  $T_\text{L}$: the energy gap does not persist for imaginary times. We can find precise conditions for the exceptional point onset, by analyzing imaginary-time fluctuations of $\dot{\phi}$, see Fig.~\ref{fig1}(e). The noise power spectrum of $\dot{\phi}_\tau$ evaluated for paths generated by the bare system is~\cite{Supplemental_Material},
\begin{equation}\label{eq_Sv_0}
    S_{\dot{\phi}}(\omega_k)=\int_0^\beta d\tau e^{i\omega_k(\tau-\tau^\prime)}\left\langle\dot{\phi}_\tau \dot{\phi}_{\tau^\prime}\right\rangle_\text{paths}=\frac{\omega_k^2}{c\omega_k^2+1/l}\ ,
\end{equation}
with Matsubara frequencies $\omega_k=2\pi k/\beta$. The high frequency tail is constant white noise, and as such, the voltage variance diverges, $\langle \dot{\phi}_\tau^2\rangle \rightarrow \infty$. However, the system contains a natural UV cutoff due to the energy gap $\gamma$, a fact that can be shown via a time-dependent basis transformation~\cite{Supplemental_Material}. This cutoff regularizes voltage fluctuations felt by the pseudo-spin to the finite value $\langle \dot{\phi}_\tau^2\rangle\sim \gamma/c$. Inserting this expectation value into the criterion for the appearance of exceptional points, $\gamma^2=\lambda^2\langle \dot{\phi}^2_\tau\rangle$, we get the critical condition $\gamma\sim \lambda^2/c$ -- coinciding with the critical value for $\gamma$, at which the system transitions from convex to nonconvex, thus connecting the superradiant transition with PI exceptional points.

We conclude that the issue is that of adiabaticity: below the phase transition, voltage fluctuations are sufficiently weak for exceptional point to be irrelevant, justifying an adiabatic approximation in Eq.~\eqref{eq_Z_PI}, replacing the operator $L$ by the eigenvalue with real part closest zero, $-c\dot{\phi}_{\tau}^2/2+\sqrt{\gamma^2-\lambda^2\dot{\phi}_\tau^2}\approx -c_\text{eff}\dot{\phi}_{\tau}^2/2$~\footnote{Via analytic continuation from imaginary times back to real times, we now understand why in Eq.~\eqref{eq_TL_low} we had to choose the higher, and not the lower eigenenergy.}. At and above the phase transition, however, PI voltage fluctuations increase to the extent that the exceptional points are reached, and adiabaticity of the paths $\phi_\tau$ is broken.

\textit{Including dissipation. --} Nonetheless, the low-energy Lagrangian in Eq.~\eqref{eq_TL_low} can be valid above the transition in a \textit{different regime}. Shunting the circuit with an Ohmic resistance $R$, see Fig.~\ref{fig1}(a),
we get for the noise power~\cite{Altland_Simons_book}
\begin{equation}\label{eq_Sv_diss}
    S_{\dot{\phi}}(\omega_k)=\frac{\omega_k^2}{c\omega_k^2+\vert\omega_k\vert/(2\pi\eta)+1/l}\ ,
\end{equation}
with the damping coefficient $1/\eta=R_Q/R$ ($R_Q$ is the resistance quantum). Still assuming $1/\beta \gg 1/l$, we get a transition from weak to strong dissipation when $1/\eta\gtrsim c\gamma$, where the voltage noise goes from $\langle \dot{\phi}_\tau^2\rangle\sim \gamma/c$ to the much smaller value $\langle \dot{\phi}_\tau^2\rangle\sim \eta\gamma^2$. Dissipation thus smoothens the paths. Adiabaticity with respect to exceptional points is guaranteed as long as $1/\eta\gtrsim \lambda^2$. Combining the nonconvexity condition $c\gamma\sim \lambda^2$, with the above strong dissipation and adiabaticity conditions, we get $1/\eta \gtrsim \lambda^2,c\gamma$, meaning that no restriction exists for the relative magnitudes of $\lambda^2$ and $c\gamma$, and the point $\lambda^2\sim c\gamma$ is crossed without infringing the exceptional points. The adiabatic, low-energy $L\approx c_\text{eff}\dot{\phi}^2/2-V(\phi)$ in Eq.~\eqref{eq_TL_low} thus retains its validity. Moreover, the point $\lambda^2\sim c\gamma$ no longer marks a phase transition. This is measurable, as for nonzero $1/\eta$, the heat capacity is sensitive to the capacitance. We find for $\beta/(\eta c_\text{eff})> 1$~\cite{Supplemental_Material}
\begin{equation}\label{eq_heat_capacity}
    C_v\approx 1+\frac{2\pi^2\eta c_\text{eff}}{\beta}\ .
\end{equation}
Instead of the dip in $C_v$ from $1$ to $3/4$ at the critical point, we now observe the aforementioned smooth crossover of the renormalized capacitance $c_\text{eff}$.

\textit{Generalization. --} The above concepts and findings can be generalized. Consider a coupling to a many-body system (referred to as 'polarizer' from here on),
\begin{equation}\label{eq_general_model}
    \widehat{T}(N)=\frac{1}{2c}\left(N+\Lambda\right)^{2}+\Gamma\ , \qquad \widehat{T}^*(\dot{\phi})=\frac{c}{2}\dot{\phi}^{2}-\dot{\phi}\Lambda-\Gamma\ ,
\end{equation}
where operator $\Lambda$ is proportional to the total polarizer dipole moment, and operator $\Gamma$ describes its intrinsic dynamics~\footnote{Both $\Lambda$ and $\Gamma$ are Hermitean operators.}. The linear coupling in $\dot{\phi}$ is generic, since for first principle modelling of electronic degrees of freedom, any externally applied voltage enters the Hamiltonian linearly. Thus, for imaginary times, $\dot{\phi}\rightarrow i\dot{\phi}$, the emergence of exceptional points is inescapable as $\dot{\phi}$ increases.

\begin{figure}
    \centering
    \includegraphics[width=0.9\linewidth]{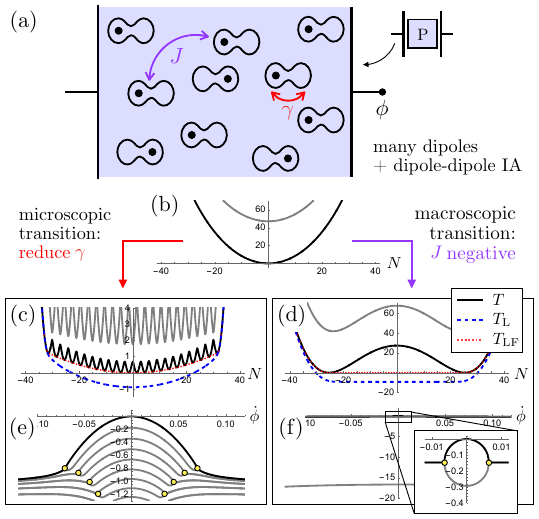}
    \caption{Generalization to a model with many dipoles (a). For sufficiently large $\gamma$ and $J>0$, the low-energy spectrum of $\widehat{T}(N)$ is convex (b). By either reducing $\gamma$, or rendering $J<0$, the model undergoes a microscopic (c), respectively macroscopic (d) phase transition, with a nonconvex ground state (black lines). Legendre- and Legendre-Fenchel-transformed kinetic energies are again drawn in blue dashed and red dotted lines in (c,d). The imaginary time eigenspectrum of $\widehat{T}^*(i\dot{\phi})$ is shown in (e,f), exhibiting exceptional points (yellow dots) correlating with the number of minima of the nonconvex $T$ from the corresponding Hamiltonian (c,d). In (b-f), the y-axis is energy, again in units of $1/c$, and $\lambda=3$, $S=21/2$. The other parameters are $c\gamma=20$ (b,d,f), $c\gamma=0.1$ (c,e), and $cJ=0.01$ (b,c,e), $cJ=-1.5$ (d,f).}
    \label{fig2}
\end{figure}

Regarding nonconvexity, there are two types of transitions. To see this, notice that the first part ($\sim 1/c$) provides as the eigenspectrum a set of parabolas as a function of $N$, shifted by the eigenvalues of $\Lambda$. A finite $\Gamma$ generically leads to hybridization, i.e., an avoided crossing of the parabolas. Consequently, for sufficiently large coupling, the spectrum is in general a convex function of $N$. This is the same type of phase transition as in the toy model presented above, now generalized for a macroscopic polarizer. We call it the 'microscopic' transition, as the nonconvex features here resolve the individual, quantized dipole moments.
The second type can be thought of as a 'macroscopic' phase transition, where the intrinsic dynamics within $\Gamma$ not only lead to hybridization, but also include a mechanism that energetically favors strongly separated, macroscopic states.

These two transition types can be illustrated by an explicit polarizer model. We take $\Lambda = \lambda S_z$ and $\Gamma=\gamma S_x+J S_z^2$, where $S_{x,z}$ are the components of a large pseudo-spin of size $S\gg 1$. Such models have been studied in the past in the context of the Dicke model~\cite{Hwang_2016,Peng_2019} and spin-squeezed states~\cite{Rojo_2003}~\footnote{As an aside: for $J>0$, if $S_z$ remains sufficiently small compared to $S$, we can approximate the dynamics as an effective Josephson effect, $S_z\sim n$, $S_x\sim\cos(\varphi)$ with $[n,\varphi]=i$. Incidentally, this results in a model that has been recently proposed to exist when coupling charge islands to charge qubits, resulting in a quasiperiodic non-linear capacitor~\cite{Herrig_2023,Herrig_2025}.}. This model can be derived from an ensemble of two-level dipole moments and remains exactly diagonalizable for large spins~\cite{Supplemental_Material}. The model and the two transition types are shown in Fig. \ref{fig2}.

We again compare the $T$ that results from the lowest eigenvalue of $\widehat{T}$, with $T_\text{L}$ (the indirect calculation via Legendre transforming $T^*$, the highest eigenvalue of $\widehat{T}^*$), and with $T_\text{LF}$ (again, a double Legendre-Fenchel-transformed $T$), see Fig.~\ref{fig2}(b,c,d). Above the transition ($\gamma$ largest energy scale, $J$ small but positive), we can describe the low-energy physics again with a well-defined effective capacitance $c_\text{eff}\approx c+S\lambda^2/\gamma$. The microscopic transition (lowering $\gamma$ below the critical value) qualitatively resembles the two-level toy model. In particular, the low-energy $T_\text{L}$ again features a smooth crossover from weakly to strongly renormalized effective capacitance, relevant again for the strongly dissipative regime, where Eq.~\eqref{eq_heat_capacity} still holds.

In contrast, the macroscopic phase transition occurs
for $J$ weakly negative (while keeping $\gamma$ large), where the system quickly reaches a tipping point, due to the competition between the $\gamma S_x$ and $JS_z^2$ terms. With their respective scaling $\sim S\gamma$ and $\sim S^2J$, we see that when $\vert J\vert > \gamma/S$, the system topples towards a state maximizing (instead of minimizing) $S_z^2$. In the thermodynamic limit $S\rightarrow \infty$ this happens for arbitrarily small, negative $J$. The Legendre-transformed $T_L$ then features an almost flat energy with respect to $N$, see Fig.~\ref{fig2}(d), representing a nearly diverging effective capacitance. This feature follows from a residual avoided crossing between the $S_z=-S$ and $S_z=+S$ states in the eigenspectrum of $\widehat{T}^*$, which for $\vert J\vert >\gamma/S$ is exponentially suppressed. If $\beta$ and $1/\eta$ are sufficiently large, we can still perform the adiabatic approximation (avoiding exceptional points) even in this extreme regime. We thus witness a first order phase transition (instead of a cross over) due to the sudden increase of the effective capacitance. 
For moderate dissipation however, the exceptional point cannot be avoided any more. Contrary to all other cases discussed above, which allowed for straightforward analytic calculations, quantitative predictions require here numerical evaluation of the path integral. Importantly though, such a calculation is still tractable (suitable for future research): while the gap between the two states involved in the exceptional point formation is (exponentially) small, there still exists a large gap to the higher states [black versus grey lines in Fig.~\ref{fig2}(f)]. This requires the evaluation of Eq.~\eqref{eq_Z_PI} truncated to a two-by-two model -- ultimately mapping back to the toy model discussed above. Regarding truncation, we generally observe that the number of minima in the Hamiltonian, Fig.~\ref{fig2}(c,d), directly correlates with the number of relevant low-energy exceptional points the imaginary time Lagrangian, Fig.~\ref{fig2}(e,f), and thus to the number of states that have to be retained. Therefore, we expect that PI calculations for moderate $1/\eta$ are numerically less (more) expensive for the macroscopic (microscopic) transition.

Finally, we comment on the usefulness of $T_\text{LF}$ for the macroscopic model. Here, it actually agrees with $T_\text{L}$ to a much higher degree (compare blue dashed and red dotted lines in Fig.~\ref{fig2}). In particular, both kinetic energy version are useful to capture phase transitions (or the lack thereof) in $c_\text{eff}$ in the strongly dissipative regime.

\textit{Conclusions and outlook. --} With explicit models, we have clarified the usefulness of various low-energy approximations on the $H$ and $L$ level, as well as their correspondence (or lack thereof) via Legendre(-Fenchel) transformations. Inconsistent low-energy $L$ and $H$ are perfectly acceptable, as they apply in complementary regimes of strong and weak dissipation, marking a dissipative phase transition. Our framework allows for efficient computation of observables (often analytically) with PI methods, where imaginary time exceptional points are linked to nonconvexity in real time. In contrast to commonly discussed dissipative phase transitions renormalizing the potential energy (e.g., the Schmid-Bulgadaev transition~\cite{Schmid_1983,Bulgadaev_1984}), we find a \textit{dual} transition modifying the kinetic energy -- not requiring RG methods, but following from simple path adiabaticity considerations. Finally, note that while above models always yielded convex Lagrangian kinetic energies, there is no fundamental principle forbidding nonconvexity in $L$, realizing classical time crystals~\cite{Shapere_2012}. Our microscopic approach could thus resolve incompatibility issues between $L$ and $H$ that are currently being debated in this field \cite{Henneaux_1987,Shapere_2012b,Choudhury_2019,Zhao_2013,Chi_2014,Dai_2020}, and naturally provide access to the quantum treatment of classical time crystals, without the need to develop new quantization methods.

\begin{acknowledgments}
We warmly thank David~P.~DiVincenzo as well as Íñigo Luis Egusquiza for fruitful discussions. This work has been funded by the German Federal Ministry of Education and Research within the funding program Photonic Research Germany under the contract number 13N14891. 
\end{acknowledgments}

\bibliographystyle{apsrev4-2}
\bibliography{references}

\begin{thebibliography}{94}%
\makeatletter
\providecommand \@ifxundefined [1]{%
 \@ifx{#1\undefined}
}%
\providecommand \@ifnum [1]{%
 \ifnum #1\expandafter \@firstoftwo
 \else \expandafter \@secondoftwo
 \fi
}%
\providecommand \@ifx [1]{%
 \ifx #1\expandafter \@firstoftwo
 \else \expandafter \@secondoftwo
 \fi
}%
\providecommand \natexlab [1]{#1}%
\providecommand \enquote  [1]{``#1''}%
\providecommand \bibnamefont  [1]{#1}%
\providecommand \bibfnamefont [1]{#1}%
\providecommand \citenamefont [1]{#1}%
\providecommand \href@noop [0]{\@secondoftwo}%
\providecommand \href [0]{\begingroup \@sanitize@url \@href}%
\providecommand \@href[1]{\@@startlink{#1}\@@href}%
\providecommand \@@href[1]{\endgroup#1\@@endlink}%
\providecommand \@sanitize@url [0]{\catcode `\\12\catcode `\$12\catcode
  `\&12\catcode `\#12\catcode `\^12\catcode `\_12\catcode `\%12\relax}%
\providecommand \@@startlink[1]{}%
\providecommand \@@endlink[0]{}%
\providecommand \url  [0]{\begingroup\@sanitize@url \@url }%
\providecommand \@url [1]{\endgroup\@href {#1}{\urlprefix }}%
\providecommand \urlprefix  [0]{URL }%
\providecommand \Eprint [0]{\href }%
\providecommand \doibase [0]{https://doi.org/}%
\providecommand \selectlanguage [0]{\@gobble}%
\providecommand \bibinfo  [0]{\@secondoftwo}%
\providecommand \bibfield  [0]{\@secondoftwo}%
\providecommand \translation [1]{[#1]}%
\providecommand \BibitemOpen [0]{}%
\providecommand \bibitemStop [0]{}%
\providecommand \bibitemNoStop [0]{.\EOS\space}%
\providecommand \EOS [0]{\spacefactor3000\relax}%
\providecommand \BibitemShut  [1]{\csname bibitem#1\endcsname}%
\let\auto@bib@innerbib\@empty
\bibitem [{\citenamefont {Schwabl}(2007)}]{Schwabl2007}%
  \BibitemOpen
  \bibfield  {author} {\bibinfo {author} {\bibfnamefont {F.}~\bibnamefont
  {Schwabl}},\ }\href@noop {} {\emph {\bibinfo {title} {Quantum Mechanics}}},\
  \bibinfo {edition} {4th}\ ed.\ (\bibinfo  {publisher} {Springer},\ \bibinfo
  {address} {Berlin, Germany},\ \bibinfo {year} {2007})\BibitemShut {NoStop}%
\bibitem [{\citenamefont {Zee}(2010)}]{Zee2010}%
  \BibitemOpen
  \bibfield  {author} {\bibinfo {author} {\bibfnamefont {A.}~\bibnamefont
  {Zee}},\ }\href@noop {} {\emph {\bibinfo {title} {Quantum field theory in a
  nutshell}}},\ \bibinfo {edition} {second edition}\ ed.\ (\bibinfo
  {publisher} {Princeton University Press},\ \bibinfo {year}
  {2010})\BibitemShut {NoStop}%
\bibitem [{Note1()}]{Note1}%
  \BibitemOpen
  \bibinfo {note} {Strictly speaking, the connection between $H$ and $L$ for PI
  is related to a Fourier transformation, and not the Legendre transformation.
  The two are only equivalent for \protect \textit {harmonic} kinetic
  energies.}\BibitemShut {Stop}%
\bibitem [{Note2()}]{Note2}%
  \BibitemOpen
  \bibinfo {note} {In line with the previous footnote, the conventional PI
  approach loses validity for \protect \textit {any} anharmonic kinetic energy,
  an issue that is well-known in the relativistic community, see, e.g.,
  Ref.~\cite {Padmanabhan_2018} and references therein.}\BibitemShut {Stop}%
\bibitem [{\citenamefont {Padmanabhan}(2018)}]{Padmanabhan_2018}%
  \BibitemOpen
  \bibfield  {author} {\bibinfo {author} {\bibfnamefont {T.}~\bibnamefont
  {Padmanabhan}},\ }\href {https://doi.org/10.1140/epjc/s10052-018-6039-y}
  {\bibfield  {journal} {\bibinfo  {journal} {The European Physical Journal C}\
  }\textbf {\bibinfo {volume} {78}},\ \bibinfo {pages} {563} (\bibinfo {year}
  {2018})}\BibitemShut {NoStop}%
\bibitem [{\citenamefont {Bergmann}\ and\ \citenamefont
  {Brunings}(1949)}]{Bergmann_1949}%
  \BibitemOpen
  \bibfield  {author} {\bibinfo {author} {\bibfnamefont {P.~G.}\ \bibnamefont
  {Bergmann}}\ and\ \bibinfo {author} {\bibfnamefont {J.~H.~M.}\ \bibnamefont
  {Brunings}},\ }\href {https://doi.org/10.1103/RevModPhys.21.480} {\bibfield
  {journal} {\bibinfo  {journal} {Rev. Mod. Phys.}\ }\textbf {\bibinfo {volume}
  {21}},\ \bibinfo {pages} {480} (\bibinfo {year} {1949})}\BibitemShut
  {NoStop}%
\bibitem [{\citenamefont {Faddeev}\ and\ \citenamefont
  {Jackiw}(1988)}]{Faddeev_1988}%
  \BibitemOpen
  \bibfield  {author} {\bibinfo {author} {\bibfnamefont {L.}~\bibnamefont
  {Faddeev}}\ and\ \bibinfo {author} {\bibfnamefont {R.}~\bibnamefont
  {Jackiw}},\ }\href {https://doi.org/10.1103/PhysRevLett.60.1692} {\bibfield
  {journal} {\bibinfo  {journal} {Phys. Rev. Lett.}\ }\textbf {\bibinfo
  {volume} {60}},\ \bibinfo {pages} {1692} (\bibinfo {year}
  {1988})}\BibitemShut {NoStop}%
\bibitem [{\citenamefont {Brown}(2022)}]{Brown_2022}%
  \BibitemOpen
  \bibfield  {author} {\bibinfo {author} {\bibfnamefont {J.~D.}\ \bibnamefont
  {Brown}},\ }\bibfield  {journal} {\bibinfo  {journal} {Universe}\ }\textbf
  {\bibinfo {volume} {8}},\ \href {https://doi.org/10.3390/universe8030171}
  {10.3390/universe8030171} (\bibinfo {year} {2022})\BibitemShut {NoStop}%
\bibitem [{\citenamefont {Parra-Rodriguez}\ and\ \citenamefont
  {Egusquiza}(2024)}]{ParraRodriguez2024}%
  \BibitemOpen
  \bibfield  {author} {\bibinfo {author} {\bibfnamefont {A.}~\bibnamefont
  {Parra-Rodriguez}}\ and\ \bibinfo {author} {\bibfnamefont {I.~L.}\
  \bibnamefont {Egusquiza}},\ }\href@noop {} {\bibfield  {journal} {\bibinfo
  {journal} {Quantum}\ }\textbf {\bibinfo {volume} {8}},\ \bibinfo {pages}
  {1466} (\bibinfo {year} {2024})}\BibitemShut {NoStop}%
\bibitem [{\citenamefont {Rymarz}\ and\ \citenamefont
  {DiVincenzo}(2023)}]{Rymarz_2023}%
  \BibitemOpen
  \bibfield  {author} {\bibinfo {author} {\bibfnamefont {M.}~\bibnamefont
  {Rymarz}}\ and\ \bibinfo {author} {\bibfnamefont {D.~P.}\ \bibnamefont
  {DiVincenzo}},\ }\href {https://doi.org/10.1103/PhysRevX.13.021017}
  {\bibfield  {journal} {\bibinfo  {journal} {Phys. Rev. X}\ }\textbf {\bibinfo
  {volume} {13}},\ \bibinfo {pages} {021017} (\bibinfo {year}
  {2023})}\BibitemShut {NoStop}%
\bibitem [{\citenamefont {Egusquiza}\ and\ \citenamefont
  {Parra-Rodriguez}(2025)}]{Egusquiza_2025comment}%
  \BibitemOpen
  \bibfield  {author} {\bibinfo {author} {\bibfnamefont {I.~L.}\ \bibnamefont
  {Egusquiza}}\ and\ \bibinfo {author} {\bibfnamefont {A.}~\bibnamefont
  {Parra-Rodriguez}},\ }\href {https://doi.org/10.1103/PhysRevX.15.028001}
  {\bibfield  {journal} {\bibinfo  {journal} {Phys. Rev. X}\ }\textbf {\bibinfo
  {volume} {15}},\ \bibinfo {pages} {028001} (\bibinfo {year}
  {2025})}\BibitemShut {NoStop}%
\bibitem [{\citenamefont {DiVincenzo}\ and\ \citenamefont
  {Rymarz}(2025)}]{DiVincenzo_2025reply}%
  \BibitemOpen
  \bibfield  {author} {\bibinfo {author} {\bibfnamefont {D.~P.}\ \bibnamefont
  {DiVincenzo}}\ and\ \bibinfo {author} {\bibfnamefont {M.}~\bibnamefont
  {Rymarz}},\ }\href {https://doi.org/10.1103/PhysRevX.15.028002} {\bibfield
  {journal} {\bibinfo  {journal} {Phys. Rev. X}\ }\textbf {\bibinfo {volume}
  {15}},\ \bibinfo {pages} {028002} (\bibinfo {year} {2025})}\BibitemShut
  {NoStop}%
\bibitem [{\citenamefont {Shapere}\ and\ \citenamefont
  {Wilczek}(2012{\natexlab{a}})}]{Shapere_2012}%
  \BibitemOpen
  \bibfield  {author} {\bibinfo {author} {\bibfnamefont {A.}~\bibnamefont
  {Shapere}}\ and\ \bibinfo {author} {\bibfnamefont {F.}~\bibnamefont
  {Wilczek}},\ }\href {https://doi.org/10.1103/PhysRevLett.109.160402}
  {\bibfield  {journal} {\bibinfo  {journal} {Phys. Rev. Lett.}\ }\textbf
  {\bibinfo {volume} {109}},\ \bibinfo {pages} {160402} (\bibinfo {year}
  {2012}{\natexlab{a}})}\BibitemShut {NoStop}%
\bibitem [{\citenamefont {Zhao}\ \emph {et~al.}(2013)\citenamefont {Zhao},
  \citenamefont {Yu},\ and\ \citenamefont {Xu}}]{Zhao_2013}%
  \BibitemOpen
  \bibfield  {author} {\bibinfo {author} {\bibfnamefont {L.}~\bibnamefont
  {Zhao}}, \bibinfo {author} {\bibfnamefont {P.}~\bibnamefont {Yu}},\ and\
  \bibinfo {author} {\bibfnamefont {W.}~\bibnamefont {Xu}},\ }\href
  {https://doi.org/10.1142/S0217732313500028} {\bibfield  {journal} {\bibinfo
  {journal} {Modern Physics Letters A}\ }\textbf {\bibinfo {volume} {28}},\
  \bibinfo {pages} {1350002} (\bibinfo {year} {2013})},\ \Eprint
  {https://arxiv.org/abs/https://doi.org/10.1142/S0217732313500028}
  {https://doi.org/10.1142/S0217732313500028} \BibitemShut {NoStop}%
\bibitem [{\citenamefont {Dai}\ \emph {et~al.}(2020)\citenamefont {Dai},
  \citenamefont {Niemi},\ and\ \citenamefont {Peng}}]{Dai_2020}%
  \BibitemOpen
  \bibfield  {author} {\bibinfo {author} {\bibfnamefont {J.}~\bibnamefont
  {Dai}}, \bibinfo {author} {\bibfnamefont {A.~J.}\ \bibnamefont {Niemi}},\
  and\ \bibinfo {author} {\bibfnamefont {X.}~\bibnamefont {Peng}},\ }\href
  {https://doi.org/10.1088/1367-2630/aba8d3} {\bibfield  {journal} {\bibinfo
  {journal} {New Journal of Physics}\ }\textbf {\bibinfo {volume} {22}},\
  \bibinfo {pages} {085006} (\bibinfo {year} {2020})}\BibitemShut {NoStop}%
\bibitem [{\citenamefont {Henneaux}\ \emph {et~al.}(1987)\citenamefont
  {Henneaux}, \citenamefont {Teitelboim},\ and\ \citenamefont
  {Zanelli}}]{Henneaux_1987}%
  \BibitemOpen
  \bibfield  {author} {\bibinfo {author} {\bibfnamefont {M.}~\bibnamefont
  {Henneaux}}, \bibinfo {author} {\bibfnamefont {C.}~\bibnamefont
  {Teitelboim}},\ and\ \bibinfo {author} {\bibfnamefont {J.}~\bibnamefont
  {Zanelli}},\ }\href {https://doi.org/10.1103/PhysRevA.36.4417} {\bibfield
  {journal} {\bibinfo  {journal} {Phys. Rev. A}\ }\textbf {\bibinfo {volume}
  {36}},\ \bibinfo {pages} {4417} (\bibinfo {year} {1987})}\BibitemShut
  {NoStop}%
\bibitem [{\citenamefont {Shapere}\ and\ \citenamefont
  {Wilczek}(2012{\natexlab{b}})}]{Shapere_2012b}%
  \BibitemOpen
  \bibfield  {author} {\bibinfo {author} {\bibfnamefont {A.}~\bibnamefont
  {Shapere}}\ and\ \bibinfo {author} {\bibfnamefont {F.}~\bibnamefont
  {Wilczek}},\ }\href {https://doi.org/10.1103/PhysRevLett.109.200402}
  {\bibfield  {journal} {\bibinfo  {journal} {Phys. Rev. Lett.}\ }\textbf
  {\bibinfo {volume} {109}},\ \bibinfo {pages} {200402} (\bibinfo {year}
  {2012}{\natexlab{b}})}\BibitemShut {NoStop}%
\bibitem [{\citenamefont {Choudhury}\ and\ \citenamefont
  {Guha}(2019)}]{Choudhury_2019}%
  \BibitemOpen
  \bibfield  {author} {\bibinfo {author} {\bibfnamefont {A.~G.}\ \bibnamefont
  {Choudhury}}\ and\ \bibinfo {author} {\bibfnamefont {P.}~\bibnamefont
  {Guha}},\ }\href {https://doi.org/10.1142/S0217732319502638} {\bibfield
  {journal} {\bibinfo  {journal} {Modern Physics Letters A}\ }\textbf {\bibinfo
  {volume} {34}},\ \bibinfo {pages} {1950263} (\bibinfo {year} {2019})},\
  \Eprint {https://arxiv.org/abs/https://doi.org/10.1142/S0217732319502638}
  {https://doi.org/10.1142/S0217732319502638} \BibitemShut {NoStop}%
\bibitem [{\citenamefont {Chi}\ and\ \citenamefont {He}(2014)}]{Chi_2014}%
  \BibitemOpen
  \bibfield  {author} {\bibinfo {author} {\bibfnamefont {H.-H.}\ \bibnamefont
  {Chi}}\ and\ \bibinfo {author} {\bibfnamefont {H.-J.}\ \bibnamefont {He}},\
  }\href {https://doi.org/https://doi.org/10.1016/j.nuclphysb.2014.05.017}
  {\bibfield  {journal} {\bibinfo  {journal} {Nuclear Physics B}\ }\textbf
  {\bibinfo {volume} {885}},\ \bibinfo {pages} {448} (\bibinfo {year}
  {2014})}\BibitemShut {NoStop}%
\bibitem [{\citenamefont {Little}(1964)}]{Little_1964}%
  \BibitemOpen
  \bibfield  {author} {\bibinfo {author} {\bibfnamefont {W.~A.}\ \bibnamefont
  {Little}},\ }\href
  {https://doi.org/https://doi.org/10.1103/PhysRev.134.A1416} {\bibfield
  {journal} {\bibinfo  {journal} {Phys. Rev.}\ }\textbf {\bibinfo {volume}
  {134}},\ \bibinfo {pages} {A1416} (\bibinfo {year} {1964})}\BibitemShut
  {NoStop}%
\bibitem [{\citenamefont {Hamo}\ \emph {et~al.}(2016)\citenamefont {Hamo},
  \citenamefont {Benyamini}, \citenamefont {Shapir}, \citenamefont {Khivrich},
  \citenamefont {Waissman}, \citenamefont {Kaasbjerg}, \citenamefont {Oreg},
  \citenamefont {von Oppen},\ and\ \citenamefont {Ilani}}]{Hamo_2016}%
  \BibitemOpen
  \bibfield  {author} {\bibinfo {author} {\bibfnamefont {A.}~\bibnamefont
  {Hamo}}, \bibinfo {author} {\bibfnamefont {A.}~\bibnamefont {Benyamini}},
  \bibinfo {author} {\bibfnamefont {I.}~\bibnamefont {Shapir}}, \bibinfo
  {author} {\bibfnamefont {I.}~\bibnamefont {Khivrich}}, \bibinfo {author}
  {\bibfnamefont {J.}~\bibnamefont {Waissman}}, \bibinfo {author}
  {\bibfnamefont {K.}~\bibnamefont {Kaasbjerg}}, \bibinfo {author}
  {\bibfnamefont {Y.}~\bibnamefont {Oreg}}, \bibinfo {author} {\bibfnamefont
  {F.}~\bibnamefont {von Oppen}},\ and\ \bibinfo {author} {\bibfnamefont
  {S.}~\bibnamefont {Ilani}},\ }\href
  {https://doi.org/https://doi.org/10.1038/nature18639} {\bibfield  {journal}
  {\bibinfo  {journal} {Nature}\ }\textbf {\bibinfo {volume} {535}},\ \bibinfo
  {pages} {395} (\bibinfo {year} {2016})}\BibitemShut {NoStop}%
\bibitem [{\citenamefont {Herrig}\ \emph {et~al.}(2023)\citenamefont {Herrig},
  \citenamefont {Pixley}, \citenamefont {K{\"o}nig},\ and\ \citenamefont
  {Riwar}}]{Herrig_2023}%
  \BibitemOpen
  \bibfield  {author} {\bibinfo {author} {\bibfnamefont {T.}~\bibnamefont
  {Herrig}}, \bibinfo {author} {\bibfnamefont {J.~H.}\ \bibnamefont {Pixley}},
  \bibinfo {author} {\bibfnamefont {E.~J.}\ \bibnamefont {K{\"o}nig}},\ and\
  \bibinfo {author} {\bibfnamefont {R.~P.}\ \bibnamefont {Riwar}},\ }\href
  {https://doi.org/10.1038/s41534-023-00786-6} {\bibfield  {journal} {\bibinfo
  {journal} {npj Quantum Information}\ }\textbf {\bibinfo {volume} {9}},\
  \bibinfo {pages} {116} (\bibinfo {year} {2023})}\BibitemShut {NoStop}%
\bibitem [{\citenamefont {Herrig}\ \emph {et~al.}(2025)\citenamefont {Herrig},
  \citenamefont {Koliofoti}, \citenamefont {Pixley}, \citenamefont {K\"onig},\
  and\ \citenamefont {Riwar}}]{Herrig_2025}%
  \BibitemOpen
  \bibfield  {author} {\bibinfo {author} {\bibfnamefont {T.}~\bibnamefont
  {Herrig}}, \bibinfo {author} {\bibfnamefont {C.}~\bibnamefont {Koliofoti}},
  \bibinfo {author} {\bibfnamefont {J.~H.}\ \bibnamefont {Pixley}}, \bibinfo
  {author} {\bibfnamefont {E.~J.}\ \bibnamefont {K\"onig}},\ and\ \bibinfo
  {author} {\bibfnamefont {R.-P.}\ \bibnamefont {Riwar}},\ }\href
  {https://doi.org/10.1103/PhysRevB.111.L201104} {\bibfield  {journal}
  {\bibinfo  {journal} {Phys. Rev. B}\ }\textbf {\bibinfo {volume} {111}},\
  \bibinfo {pages} {L201104} (\bibinfo {year} {2025})}\BibitemShut {NoStop}%
\bibitem [{\citenamefont {Landauer}(1976)}]{Landauer_1976}%
  \BibitemOpen
  \bibfield  {author} {\bibinfo {author} {\bibfnamefont {R.}~\bibnamefont
  {Landauer}},\ }\href@noop {} {\bibfield  {journal} {\bibinfo  {journal}
  {Collect. Phenom.}\ }\textbf {\bibinfo {volume} {2}},\ \bibinfo {pages} {167}
  (\bibinfo {year} {1976})}\BibitemShut {NoStop}%
\bibitem [{\citenamefont {Catalan}\ \emph {et~al.}(2015)\citenamefont
  {Catalan}, \citenamefont {Jim{\'e}nez},\ and\ \citenamefont
  {Gruverman}}]{Catalan_2015}%
  \BibitemOpen
  \bibfield  {author} {\bibinfo {author} {\bibfnamefont {G.}~\bibnamefont
  {Catalan}}, \bibinfo {author} {\bibfnamefont {D.}~\bibnamefont
  {Jim{\'e}nez}},\ and\ \bibinfo {author} {\bibfnamefont {A.}~\bibnamefont
  {Gruverman}},\ }\href {https://doi.org/10.1038/nmat4195} {\bibfield
  {journal} {\bibinfo  {journal} {Nature Materials}\ }\textbf {\bibinfo
  {volume} {14}},\ \bibinfo {pages} {137} (\bibinfo {year} {2015})}\BibitemShut
  {NoStop}%
\bibitem [{\citenamefont {Ng}\ \emph {et~al.}(2017)\citenamefont {Ng},
  \citenamefont {Hillenius},\ and\ \citenamefont {Gruverman}}]{Ng_2017}%
  \BibitemOpen
  \bibfield  {author} {\bibinfo {author} {\bibfnamefont {K.}~\bibnamefont
  {Ng}}, \bibinfo {author} {\bibfnamefont {S.~J.}\ \bibnamefont {Hillenius}},\
  and\ \bibinfo {author} {\bibfnamefont {A.}~\bibnamefont {Gruverman}},\ }\href
  {https://doi.org/https://doi.org/10.1016/j.ssc.2017.07.020} {\bibfield
  {journal} {\bibinfo  {journal} {Solid State Communications}\ }\textbf
  {\bibinfo {volume} {265}},\ \bibinfo {pages} {12 } (\bibinfo {year}
  {2017})}\BibitemShut {NoStop}%
\bibitem [{\citenamefont {Hoffmann}\ \emph {et~al.}(2018)\citenamefont
  {Hoffmann}, \citenamefont {Khan}, \citenamefont {Serrao}, \citenamefont {Lu},
  \citenamefont {Salahuddin}, \citenamefont {Pe\v{s}i\'{c}}, \citenamefont
  {Slesazeck}, \citenamefont {Schroeder},\ and\ \citenamefont
  {Mikolajick}}]{Hoffmann_2018}%
  \BibitemOpen
  \bibfield  {author} {\bibinfo {author} {\bibfnamefont {M.}~\bibnamefont
  {Hoffmann}}, \bibinfo {author} {\bibfnamefont {A.~I.}\ \bibnamefont {Khan}},
  \bibinfo {author} {\bibfnamefont {C.}~\bibnamefont {Serrao}}, \bibinfo
  {author} {\bibfnamefont {Z.}~\bibnamefont {Lu}}, \bibinfo {author}
  {\bibfnamefont {S.}~\bibnamefont {Salahuddin}}, \bibinfo {author}
  {\bibfnamefont {M.}~\bibnamefont {Pe\v{s}i\'{c}}}, \bibinfo {author}
  {\bibfnamefont {S.}~\bibnamefont {Slesazeck}}, \bibinfo {author}
  {\bibfnamefont {U.}~\bibnamefont {Schroeder}},\ and\ \bibinfo {author}
  {\bibfnamefont {T.}~\bibnamefont {Mikolajick}},\ }\href
  {https://doi.org/https://doi.org/10.1063/1.5030072} {\bibfield  {journal}
  {\bibinfo  {journal} {J. Appl. Phys.}\ }\textbf {\bibinfo {volume} {123}},\
  \bibinfo {pages} {184101} (\bibinfo {year} {2018})}\BibitemShut {NoStop}%
\bibitem [{\citenamefont {Luk'yanchuk}\ \emph {et~al.}(2019)\citenamefont
  {Luk'yanchuk}, \citenamefont {Tikhonov}, \citenamefont {Sen{\'e}},
  \citenamefont {Razumnaya},\ and\ \citenamefont {Vinokur}}]{Lukyanchuk_2019}%
  \BibitemOpen
  \bibfield  {author} {\bibinfo {author} {\bibfnamefont {I.}~\bibnamefont
  {Luk'yanchuk}}, \bibinfo {author} {\bibfnamefont {Y.}~\bibnamefont
  {Tikhonov}}, \bibinfo {author} {\bibfnamefont {A.}~\bibnamefont {Sen{\'e}}},
  \bibinfo {author} {\bibfnamefont {A.}~\bibnamefont {Razumnaya}},\ and\
  \bibinfo {author} {\bibfnamefont {V.~M.}\ \bibnamefont {Vinokur}},\ }\href
  {https://doi.org/https://doi.org/10.1038/s42005-019-0121-0} {\bibfield
  {journal} {\bibinfo  {journal} {Commun. Phys.}\ }\textbf {\bibinfo {volume}
  {2}},\ \bibinfo {pages} {22} (\bibinfo {year} {2019})}\BibitemShut {NoStop}%
\bibitem [{\citenamefont {Hoffmann}\ \emph {et~al.}(2020)\citenamefont
  {Hoffmann}, \citenamefont {Slesazeck}, \citenamefont {Schroeder},\ and\
  \citenamefont {Mikolajick}}]{Hoffmann_2020}%
  \BibitemOpen
  \bibfield  {author} {\bibinfo {author} {\bibfnamefont {M.}~\bibnamefont
  {Hoffmann}}, \bibinfo {author} {\bibfnamefont {S.}~\bibnamefont {Slesazeck}},
  \bibinfo {author} {\bibfnamefont {U.}~\bibnamefont {Schroeder}},\ and\
  \bibinfo {author} {\bibfnamefont {T.}~\bibnamefont {Mikolajick}},\ }\href
  {https://doi.org/10.1038/s41928-020-00474-9} {\bibfield  {journal} {\bibinfo
  {journal} {Nature Electronics}\ }\textbf {\bibinfo {volume} {3}},\ \bibinfo
  {pages} {504} (\bibinfo {year} {2020})}\BibitemShut {NoStop}%
\bibitem [{\citenamefont {Giordano}(1988)}]{Giordano1988}%
  \BibitemOpen
  \bibfield  {author} {\bibinfo {author} {\bibfnamefont {N.}~\bibnamefont
  {Giordano}},\ }\href {https://doi.org/10.1103/PhysRevLett.61.2137} {\bibfield
   {journal} {\bibinfo  {journal} {Phys. Rev. Lett.}\ }\textbf {\bibinfo
  {volume} {61}},\ \bibinfo {pages} {2137} (\bibinfo {year}
  {1988})}\BibitemShut {NoStop}%
\bibitem [{\citenamefont {Bezryadin}\ \emph {et~al.}(2000)\citenamefont
  {Bezryadin}, \citenamefont {Lau},\ and\ \citenamefont
  {Tinkham}}]{Bezryadin2000}%
  \BibitemOpen
  \bibfield  {author} {\bibinfo {author} {\bibfnamefont {A.}~\bibnamefont
  {Bezryadin}}, \bibinfo {author} {\bibfnamefont {C.}~\bibnamefont {Lau}},\
  and\ \bibinfo {author} {\bibfnamefont {M.}~\bibnamefont {Tinkham}},\ }\href
  {https://doi.org/10.1038/35010060} {\bibfield  {journal} {\bibinfo  {journal}
  {Nature}\ }\textbf {\bibinfo {volume} {404}},\ \bibinfo {pages} {971}
  (\bibinfo {year} {2000})}\BibitemShut {NoStop}%
\bibitem [{\citenamefont {Lau}\ \emph {et~al.}(2001)\citenamefont {Lau},
  \citenamefont {Markovic}, \citenamefont {Bockrath}, \citenamefont
  {Bezryadin},\ and\ \citenamefont {Tinkham}}]{Lau_2001}%
  \BibitemOpen
  \bibfield  {author} {\bibinfo {author} {\bibfnamefont {C.~N.}\ \bibnamefont
  {Lau}}, \bibinfo {author} {\bibfnamefont {N.}~\bibnamefont {Markovic}},
  \bibinfo {author} {\bibfnamefont {M.}~\bibnamefont {Bockrath}}, \bibinfo
  {author} {\bibfnamefont {A.}~\bibnamefont {Bezryadin}},\ and\ \bibinfo
  {author} {\bibfnamefont {M.}~\bibnamefont {Tinkham}},\ }\href
  {https://doi.org/10.1103/PhysRevLett.87.217003} {\bibfield  {journal}
  {\bibinfo  {journal} {Phys. Rev. Lett.}\ }\textbf {\bibinfo {volume} {87}},\
  \bibinfo {pages} {217003} (\bibinfo {year} {2001})}\BibitemShut {NoStop}%
\bibitem [{\citenamefont {B\"uchler}\ \emph {et~al.}(2004)\citenamefont
  {B\"uchler}, \citenamefont {Geshkenbein},\ and\ \citenamefont
  {Blatter}}]{Buchler2004}%
  \BibitemOpen
  \bibfield  {author} {\bibinfo {author} {\bibfnamefont {H.~P.}\ \bibnamefont
  {B\"uchler}}, \bibinfo {author} {\bibfnamefont {V.~B.}\ \bibnamefont
  {Geshkenbein}},\ and\ \bibinfo {author} {\bibfnamefont {G.}~\bibnamefont
  {Blatter}},\ }\href {https://doi.org/10.1103/PhysRevLett.92.067007}
  {\bibfield  {journal} {\bibinfo  {journal} {Phys. Rev. Lett.}\ }\textbf
  {\bibinfo {volume} {92}},\ \bibinfo {pages} {067007} (\bibinfo {year}
  {2004})}\BibitemShut {NoStop}%
\bibitem [{\citenamefont {Mooij}\ and\ \citenamefont
  {Nazarov}(2006)}]{Mooij_2006}%
  \BibitemOpen
  \bibfield  {author} {\bibinfo {author} {\bibfnamefont {J.~E.}\ \bibnamefont
  {Mooij}}\ and\ \bibinfo {author} {\bibfnamefont {Y.~V.}\ \bibnamefont
  {Nazarov}},\ }\href {https://doi.org/10.1038/nphys234} {\bibfield  {journal}
  {\bibinfo  {journal} {Nature Physics}\ }\textbf {\bibinfo {volume} {2}},\
  \bibinfo {pages} {169} (\bibinfo {year} {2006})}\BibitemShut {NoStop}%
\bibitem [{\citenamefont {Astafiev}\ \emph {et~al.}(2012)\citenamefont
  {Astafiev}, \citenamefont {Ioffe}, \citenamefont {Kafanov}, \citenamefont
  {Pashkin}, \citenamefont {Arutyunov}, \citenamefont {Shahar}, \citenamefont
  {Cohen},\ and\ \citenamefont {Tsai}}]{Astafiev_2012}%
  \BibitemOpen
  \bibfield  {author} {\bibinfo {author} {\bibfnamefont {O.~V.}\ \bibnamefont
  {Astafiev}}, \bibinfo {author} {\bibfnamefont {L.~B.}\ \bibnamefont {Ioffe}},
  \bibinfo {author} {\bibfnamefont {S.}~\bibnamefont {Kafanov}}, \bibinfo
  {author} {\bibfnamefont {Y.~A.}\ \bibnamefont {Pashkin}}, \bibinfo {author}
  {\bibfnamefont {K.~Y.}\ \bibnamefont {Arutyunov}}, \bibinfo {author}
  {\bibfnamefont {D.}~\bibnamefont {Shahar}}, \bibinfo {author} {\bibfnamefont
  {O.}~\bibnamefont {Cohen}},\ and\ \bibinfo {author} {\bibfnamefont {J.~S.}\
  \bibnamefont {Tsai}},\ }\href {https://doi.org/10.1038/nature10930}
  {\bibfield  {journal} {\bibinfo  {journal} {Nature}\ }\textbf {\bibinfo
  {volume} {484}},\ \bibinfo {pages} {355} (\bibinfo {year}
  {2012})}\BibitemShut {NoStop}%
\bibitem [{\citenamefont {de~Graaf}\ \emph {et~al.}(2018)\citenamefont
  {de~Graaf}, \citenamefont {Skacel}, \citenamefont {H{\"o}nigl-Decrinis},
  \citenamefont {Shaikhaidarov}, \citenamefont {Rotzinger}, \citenamefont
  {Linzen}, \citenamefont {Ziegler}, \citenamefont {H{\"u}bner}, \citenamefont
  {Meyer}, \citenamefont {Antonov}, \citenamefont {Il'ichev}, \citenamefont
  {Ustinov}, \citenamefont {Tzalenchuk},\ and\ \citenamefont
  {Astafiev}}]{deGraaf_2018}%
  \BibitemOpen
  \bibfield  {author} {\bibinfo {author} {\bibfnamefont {S.~E.}\ \bibnamefont
  {de~Graaf}}, \bibinfo {author} {\bibfnamefont {S.~T.}\ \bibnamefont
  {Skacel}}, \bibinfo {author} {\bibfnamefont {T.}~\bibnamefont
  {H{\"o}nigl-Decrinis}}, \bibinfo {author} {\bibfnamefont {R.}~\bibnamefont
  {Shaikhaidarov}}, \bibinfo {author} {\bibfnamefont {H.}~\bibnamefont
  {Rotzinger}}, \bibinfo {author} {\bibfnamefont {S.}~\bibnamefont {Linzen}},
  \bibinfo {author} {\bibfnamefont {M.}~\bibnamefont {Ziegler}}, \bibinfo
  {author} {\bibfnamefont {U.}~\bibnamefont {H{\"u}bner}}, \bibinfo {author}
  {\bibfnamefont {H.~G.}\ \bibnamefont {Meyer}}, \bibinfo {author}
  {\bibfnamefont {V.}~\bibnamefont {Antonov}}, \bibinfo {author} {\bibfnamefont
  {E.}~\bibnamefont {Il'ichev}}, \bibinfo {author} {\bibfnamefont {A.~V.}\
  \bibnamefont {Ustinov}}, \bibinfo {author} {\bibfnamefont {A.~Y.}\
  \bibnamefont {Tzalenchuk}},\ and\ \bibinfo {author} {\bibfnamefont {O.~V.}\
  \bibnamefont {Astafiev}},\ }\href {https://doi.org/10.1038/s41567-018-0097-9}
  {\bibfield  {journal} {\bibinfo  {journal} {Nature Physics}\ }\textbf
  {\bibinfo {volume} {14}},\ \bibinfo {pages} {590} (\bibinfo {year}
  {2018})}\BibitemShut {NoStop}%
\bibitem [{\citenamefont {Shaikhaidarov}\ \emph {et~al.}(2022)\citenamefont
  {Shaikhaidarov}, \citenamefont {Kim}, \citenamefont {Dunstan}, \citenamefont
  {Antonov}, \citenamefont {Linzen}, \citenamefont {Ziegler}, \citenamefont
  {Golubev}, \citenamefont {Antonov}, \citenamefont {Il'ichev},\ and\
  \citenamefont {Astafiev}}]{Shaikhaidarov_2022}%
  \BibitemOpen
  \bibfield  {author} {\bibinfo {author} {\bibfnamefont {R.~S.}\ \bibnamefont
  {Shaikhaidarov}}, \bibinfo {author} {\bibfnamefont {K.~H.}\ \bibnamefont
  {Kim}}, \bibinfo {author} {\bibfnamefont {J.~W.}\ \bibnamefont {Dunstan}},
  \bibinfo {author} {\bibfnamefont {I.~V.}\ \bibnamefont {Antonov}}, \bibinfo
  {author} {\bibfnamefont {S.}~\bibnamefont {Linzen}}, \bibinfo {author}
  {\bibfnamefont {M.}~\bibnamefont {Ziegler}}, \bibinfo {author} {\bibfnamefont
  {D.~S.}\ \bibnamefont {Golubev}}, \bibinfo {author} {\bibfnamefont {V.~N.}\
  \bibnamefont {Antonov}}, \bibinfo {author} {\bibfnamefont {E.~V.}\
  \bibnamefont {Il'ichev}},\ and\ \bibinfo {author} {\bibfnamefont {O.~V.}\
  \bibnamefont {Astafiev}},\ }\href
  {https://doi.org/10.1038/s41586-022-04947-z} {\bibfield  {journal} {\bibinfo
  {journal} {Nature}\ }\textbf {\bibinfo {volume} {608}},\ \bibinfo {pages}
  {45} (\bibinfo {year} {2022})}\BibitemShut {NoStop}%
\bibitem [{\citenamefont {Koliofoti}\ and\ \citenamefont
  {Riwar}(2023)}]{Koliofoti_2023}%
  \BibitemOpen
  \bibfield  {author} {\bibinfo {author} {\bibfnamefont {C.}~\bibnamefont
  {Koliofoti}}\ and\ \bibinfo {author} {\bibfnamefont {R.-P.}\ \bibnamefont
  {Riwar}},\ }\href {https://doi.org/10.1038/s41534-023-00790-w} {\bibfield
  {journal} {\bibinfo  {journal} {npj Quantum Information}\ }\textbf {\bibinfo
  {volume} {9}},\ \bibinfo {pages} {125} (\bibinfo {year} {2023})}\BibitemShut
  {NoStop}%
\bibitem [{\citenamefont {Burkard}\ \emph {et~al.}(2004)\citenamefont
  {Burkard}, \citenamefont {Koch},\ and\ \citenamefont
  {DiVincenzo}}]{Burkard_2004}%
  \BibitemOpen
  \bibfield  {author} {\bibinfo {author} {\bibfnamefont {G.}~\bibnamefont
  {Burkard}}, \bibinfo {author} {\bibfnamefont {R.~H.}\ \bibnamefont {Koch}},\
  and\ \bibinfo {author} {\bibfnamefont {D.~P.}\ \bibnamefont {DiVincenzo}},\
  }\href {https://doi.org/10.1103/PhysRevB.69.064503} {\bibfield  {journal}
  {\bibinfo  {journal} {Phys. Rev. B}\ }\textbf {\bibinfo {volume} {69}},\
  \bibinfo {pages} {064503} (\bibinfo {year} {2004})}\BibitemShut {NoStop}%
\bibitem [{\citenamefont {Vool}\ and\ \citenamefont
  {Devoret}(2017)}]{Vool_2017}%
  \BibitemOpen
  \bibfield  {author} {\bibinfo {author} {\bibfnamefont {U.}~\bibnamefont
  {Vool}}\ and\ \bibinfo {author} {\bibfnamefont {M.}~\bibnamefont {Devoret}},\
  }\href {https://doi.org/https://doi.org/10.1002/cta.2359} {\bibfield
  {journal} {\bibinfo  {journal} {International Journal of Circuit Theory and
  Applications}\ }\textbf {\bibinfo {volume} {45}},\ \bibinfo {pages} {897}
  (\bibinfo {year} {2017})}\BibitemShut {NoStop}%
\bibitem [{\citenamefont {Riwar}\ and\ \citenamefont
  {DiVincenzo}(2022)}]{Riwar_2022}%
  \BibitemOpen
  \bibfield  {author} {\bibinfo {author} {\bibfnamefont {R.~P.}\ \bibnamefont
  {Riwar}}\ and\ \bibinfo {author} {\bibfnamefont {D.~P.}\ \bibnamefont
  {DiVincenzo}},\ }\href {https://doi.org/10.1038/s41534-022-00539-x}
  {\bibfield  {journal} {\bibinfo  {journal} {npj Quantum Information}\
  }\textbf {\bibinfo {volume} {8}},\ \bibinfo {pages} {36} (\bibinfo {year}
  {2022})}\BibitemShut {NoStop}%
\bibitem [{\citenamefont {Arute}\ \emph {et~al.}(2019)\citenamefont {Arute},
  \citenamefont {Arya}, \citenamefont {Babbush}, \citenamefont {Bacon},
  \citenamefont {Bardin}, \citenamefont {Barends}, \citenamefont {Biswas},
  \citenamefont {Boixo}, \citenamefont {Brandao}, \citenamefont {Buell},
  \citenamefont {Burkett}, \citenamefont {Chen}, \citenamefont {Chen},
  \citenamefont {Chiaro}, \citenamefont {Collins}, \citenamefont {Courtney},
  \citenamefont {Dunsworth}, \citenamefont {Farhi}, \citenamefont {Foxen},
  \citenamefont {Fowler}, \citenamefont {Gidney}, \citenamefont {Giustina},
  \citenamefont {Graff}, \citenamefont {Guerin}, \citenamefont {Habegger},
  \citenamefont {Harrigan}, \citenamefont {Hartmann}, \citenamefont {Ho},
  \citenamefont {Hoffmann}, \citenamefont {Huang}, \citenamefont {Humble},
  \citenamefont {Isakov}, \citenamefont {Jeffrey}, \citenamefont {Jiang},
  \citenamefont {Kafri}, \citenamefont {Kechedzhi}, \citenamefont {Kelly},
  \citenamefont {Klimov}, \citenamefont {Knysh}, \citenamefont {Korotkov},
  \citenamefont {Kostritsa}, \citenamefont {Landhuis}, \citenamefont
  {Lindmark}, \citenamefont {Lucero}, \citenamefont {Lyakh}, \citenamefont
  {Mandr{\`a}}, \citenamefont {McClean}, \citenamefont {McEwen}, \citenamefont
  {Megrant}, \citenamefont {Mi}, \citenamefont {Michielsen}, \citenamefont
  {Mohseni}, \citenamefont {Mutus}, \citenamefont {Naaman}, \citenamefont
  {Neeley}, \citenamefont {Neill}, \citenamefont {Niu}, \citenamefont {Ostby},
  \citenamefont {Petukhov}, \citenamefont {Platt}, \citenamefont {Quintana},
  \citenamefont {Rieffel}, \citenamefont {Roushan}, \citenamefont {Rubin},
  \citenamefont {Sank}, \citenamefont {Satzinger}, \citenamefont {Smelyanskiy},
  \citenamefont {Sung}, \citenamefont {Trevithick}, \citenamefont
  {Vainsencher}, \citenamefont {Villalonga}, \citenamefont {White},
  \citenamefont {Yao}, \citenamefont {Yeh}, \citenamefont {Zalcman},
  \citenamefont {Neven},\ and\ \citenamefont {Martinis}}]{Arute_2019}%
  \BibitemOpen
  \bibfield  {author} {\bibinfo {author} {\bibfnamefont {F.}~\bibnamefont
  {Arute}}, \bibinfo {author} {\bibfnamefont {K.}~\bibnamefont {Arya}},
  \bibinfo {author} {\bibfnamefont {R.}~\bibnamefont {Babbush}}, \bibinfo
  {author} {\bibfnamefont {D.}~\bibnamefont {Bacon}}, \bibinfo {author}
  {\bibfnamefont {J.~C.}\ \bibnamefont {Bardin}}, \bibinfo {author}
  {\bibfnamefont {R.}~\bibnamefont {Barends}}, \bibinfo {author} {\bibfnamefont
  {R.}~\bibnamefont {Biswas}}, \bibinfo {author} {\bibfnamefont
  {S.}~\bibnamefont {Boixo}}, \bibinfo {author} {\bibfnamefont {F.~G. S.~L.}\
  \bibnamefont {Brandao}}, \bibinfo {author} {\bibfnamefont {D.~A.}\
  \bibnamefont {Buell}}, \bibinfo {author} {\bibfnamefont {B.}~\bibnamefont
  {Burkett}}, \bibinfo {author} {\bibfnamefont {Y.}~\bibnamefont {Chen}},
  \bibinfo {author} {\bibfnamefont {Z.}~\bibnamefont {Chen}}, \bibinfo {author}
  {\bibfnamefont {B.}~\bibnamefont {Chiaro}}, \bibinfo {author} {\bibfnamefont
  {R.}~\bibnamefont {Collins}}, \bibinfo {author} {\bibfnamefont
  {W.}~\bibnamefont {Courtney}}, \bibinfo {author} {\bibfnamefont
  {A.}~\bibnamefont {Dunsworth}}, \bibinfo {author} {\bibfnamefont
  {E.}~\bibnamefont {Farhi}}, \bibinfo {author} {\bibfnamefont
  {B.}~\bibnamefont {Foxen}}, \bibinfo {author} {\bibfnamefont
  {A.}~\bibnamefont {Fowler}}, \bibinfo {author} {\bibfnamefont
  {C.}~\bibnamefont {Gidney}}, \bibinfo {author} {\bibfnamefont
  {M.}~\bibnamefont {Giustina}}, \bibinfo {author} {\bibfnamefont
  {R.}~\bibnamefont {Graff}}, \bibinfo {author} {\bibfnamefont
  {K.}~\bibnamefont {Guerin}}, \bibinfo {author} {\bibfnamefont
  {S.}~\bibnamefont {Habegger}}, \bibinfo {author} {\bibfnamefont {M.~P.}\
  \bibnamefont {Harrigan}}, \bibinfo {author} {\bibfnamefont {M.~J.}\
  \bibnamefont {Hartmann}}, \bibinfo {author} {\bibfnamefont {A.}~\bibnamefont
  {Ho}}, \bibinfo {author} {\bibfnamefont {M.}~\bibnamefont {Hoffmann}},
  \bibinfo {author} {\bibfnamefont {T.}~\bibnamefont {Huang}}, \bibinfo
  {author} {\bibfnamefont {T.~S.}\ \bibnamefont {Humble}}, \bibinfo {author}
  {\bibfnamefont {S.~V.}\ \bibnamefont {Isakov}}, \bibinfo {author}
  {\bibfnamefont {E.}~\bibnamefont {Jeffrey}}, \bibinfo {author} {\bibfnamefont
  {Z.}~\bibnamefont {Jiang}}, \bibinfo {author} {\bibfnamefont
  {D.}~\bibnamefont {Kafri}}, \bibinfo {author} {\bibfnamefont
  {K.}~\bibnamefont {Kechedzhi}}, \bibinfo {author} {\bibfnamefont
  {J.}~\bibnamefont {Kelly}}, \bibinfo {author} {\bibfnamefont {P.~V.}\
  \bibnamefont {Klimov}}, \bibinfo {author} {\bibfnamefont {S.}~\bibnamefont
  {Knysh}}, \bibinfo {author} {\bibfnamefont {A.}~\bibnamefont {Korotkov}},
  \bibinfo {author} {\bibfnamefont {F.}~\bibnamefont {Kostritsa}}, \bibinfo
  {author} {\bibfnamefont {D.}~\bibnamefont {Landhuis}}, \bibinfo {author}
  {\bibfnamefont {M.}~\bibnamefont {Lindmark}}, \bibinfo {author}
  {\bibfnamefont {E.}~\bibnamefont {Lucero}}, \bibinfo {author} {\bibfnamefont
  {D.}~\bibnamefont {Lyakh}}, \bibinfo {author} {\bibfnamefont
  {S.}~\bibnamefont {Mandr{\`a}}}, \bibinfo {author} {\bibfnamefont {J.~R.}\
  \bibnamefont {McClean}}, \bibinfo {author} {\bibfnamefont {M.}~\bibnamefont
  {McEwen}}, \bibinfo {author} {\bibfnamefont {A.}~\bibnamefont {Megrant}},
  \bibinfo {author} {\bibfnamefont {X.}~\bibnamefont {Mi}}, \bibinfo {author}
  {\bibfnamefont {K.}~\bibnamefont {Michielsen}}, \bibinfo {author}
  {\bibfnamefont {M.}~\bibnamefont {Mohseni}}, \bibinfo {author} {\bibfnamefont
  {J.}~\bibnamefont {Mutus}}, \bibinfo {author} {\bibfnamefont
  {O.}~\bibnamefont {Naaman}}, \bibinfo {author} {\bibfnamefont
  {M.}~\bibnamefont {Neeley}}, \bibinfo {author} {\bibfnamefont
  {C.}~\bibnamefont {Neill}}, \bibinfo {author} {\bibfnamefont {M.~Y.}\
  \bibnamefont {Niu}}, \bibinfo {author} {\bibfnamefont {E.}~\bibnamefont
  {Ostby}}, \bibinfo {author} {\bibfnamefont {A.}~\bibnamefont {Petukhov}},
  \bibinfo {author} {\bibfnamefont {J.~C.}\ \bibnamefont {Platt}}, \bibinfo
  {author} {\bibfnamefont {C.}~\bibnamefont {Quintana}}, \bibinfo {author}
  {\bibfnamefont {E.~G.}\ \bibnamefont {Rieffel}}, \bibinfo {author}
  {\bibfnamefont {P.}~\bibnamefont {Roushan}}, \bibinfo {author} {\bibfnamefont
  {N.~C.}\ \bibnamefont {Rubin}}, \bibinfo {author} {\bibfnamefont
  {D.}~\bibnamefont {Sank}}, \bibinfo {author} {\bibfnamefont {K.~J.}\
  \bibnamefont {Satzinger}}, \bibinfo {author} {\bibfnamefont {V.}~\bibnamefont
  {Smelyanskiy}}, \bibinfo {author} {\bibfnamefont {K.~J.}\ \bibnamefont
  {Sung}}, \bibinfo {author} {\bibfnamefont {M.~D.}\ \bibnamefont
  {Trevithick}}, \bibinfo {author} {\bibfnamefont {A.}~\bibnamefont
  {Vainsencher}}, \bibinfo {author} {\bibfnamefont {B.}~\bibnamefont
  {Villalonga}}, \bibinfo {author} {\bibfnamefont {T.}~\bibnamefont {White}},
  \bibinfo {author} {\bibfnamefont {Z.~J.}\ \bibnamefont {Yao}}, \bibinfo
  {author} {\bibfnamefont {P.}~\bibnamefont {Yeh}}, \bibinfo {author}
  {\bibfnamefont {A.}~\bibnamefont {Zalcman}}, \bibinfo {author} {\bibfnamefont
  {H.}~\bibnamefont {Neven}},\ and\ \bibinfo {author} {\bibfnamefont {J.~M.}\
  \bibnamefont {Martinis}},\ }\href {https://doi.org/10.1038/s41586-019-1666-5}
  {\bibfield  {journal} {\bibinfo  {journal} {Nature}\ }\textbf {\bibinfo
  {volume} {574}},\ \bibinfo {pages} {505} (\bibinfo {year}
  {2019})}\BibitemShut {NoStop}%
\bibitem [{IBM(2020)}]{IBM_roadmap}%
  \BibitemOpen
  \href@noop {} {}\bibinfo {howpublished} {See
  \href{https://www.ibm.com/blogs/research/2020/09/ibm-quantum-roadmap/}{https://www.ibm.com/blogs/research/2020/09/ibm-quantum-roadmap/}}
  (\bibinfo {year} {2020})\BibitemShut {NoStop}%
\bibitem [{\citenamefont {Altland}\ and\ \citenamefont
  {Simons}(2010)}]{Altland_Simons_book}%
  \BibitemOpen
  \bibfield  {author} {\bibinfo {author} {\bibfnamefont {A.}~\bibnamefont
  {Altland}}\ and\ \bibinfo {author} {\bibfnamefont {B.~D.}\ \bibnamefont
  {Simons}},\ }\href@noop {} {\emph {\bibinfo {title} {Condensed Matter Field
  Theory}}}\ (\bibinfo  {publisher} {Cambridge University Press},\ \bibinfo
  {address} {Cambridge, UK},\ \bibinfo {year} {2010})\BibitemShut {NoStop}%
\bibitem [{Note3()}]{Note3}%
  \BibitemOpen
  \bibinfo {note} {Specifically, the Lagrangian obtained from the
  Legendre-Fenchel-transformed Hamiltonian becomes in general non-smooth, and a
  second application does not recover the original nonconvex Hamiltonian, but a
  deformed, convexificated variant.}\BibitemShut {Stop}%
\bibitem [{\citenamefont {Ulrich}\ and\ \citenamefont
  {Hassler}(2016)}]{Ulrich_2016}%
  \BibitemOpen
  \bibfield  {author} {\bibinfo {author} {\bibfnamefont {J.}~\bibnamefont
  {Ulrich}}\ and\ \bibinfo {author} {\bibfnamefont {F.}~\bibnamefont
  {Hassler}},\ }\href {https://doi.org/10.1103/PhysRevB.94.094505} {\bibfield
  {journal} {\bibinfo  {journal} {Phys. Rev. B}\ }\textbf {\bibinfo {volume}
  {94}},\ \bibinfo {pages} {094505} (\bibinfo {year} {2016})}\BibitemShut
  {NoStop}%
\bibitem [{\citenamefont {Ashhab}\ and\ \citenamefont
  {Nori}(2010)}]{Ashhab_2010}%
  \BibitemOpen
  \bibfield  {author} {\bibinfo {author} {\bibfnamefont {S.}~\bibnamefont
  {Ashhab}}\ and\ \bibinfo {author} {\bibfnamefont {F.}~\bibnamefont {Nori}},\
  }\href@noop {} {\bibfield  {journal} {\bibinfo  {journal} {Phys. Rev. A}\
  }\textbf {\bibinfo {volume} {81}},\ \bibinfo {pages} {042311} (\bibinfo
  {year} {2010})}\BibitemShut {NoStop}%
\bibitem [{\citenamefont {Ashhab}(2013)}]{Ashhab_2013}%
  \BibitemOpen
  \bibfield  {author} {\bibinfo {author} {\bibfnamefont {S.}~\bibnamefont
  {Ashhab}},\ }\href {https://doi.org/10.1103/PhysRevA.87.013826} {\bibfield
  {journal} {\bibinfo  {journal} {Phys. Rev. A}\ }\textbf {\bibinfo {volume}
  {87}},\ \bibinfo {pages} {013826} (\bibinfo {year} {2013})}\BibitemShut
  {NoStop}%
\bibitem [{\citenamefont {Hwang}\ \emph {et~al.}(2015)\citenamefont {Hwang},
  \citenamefont {Puebla},\ and\ \citenamefont {Plenio}}]{Hwang_2015}%
  \BibitemOpen
  \bibfield  {author} {\bibinfo {author} {\bibfnamefont {M.-J.}\ \bibnamefont
  {Hwang}}, \bibinfo {author} {\bibfnamefont {R.}~\bibnamefont {Puebla}},\ and\
  \bibinfo {author} {\bibfnamefont {M.~B.}\ \bibnamefont {Plenio}},\
  }\href@noop {} {\bibfield  {journal} {\bibinfo  {journal} {Phys. Rev. Lett.}\
  }\textbf {\bibinfo {volume} {115}},\ \bibinfo {pages} {180404} (\bibinfo
  {year} {2015})}\BibitemShut {NoStop}%
\bibitem [{\citenamefont {Hwang}\ and\ \citenamefont
  {Plenio}(2016)}]{Hwang_2016}%
  \BibitemOpen
  \bibfield  {author} {\bibinfo {author} {\bibfnamefont {M.-J.}\ \bibnamefont
  {Hwang}}\ and\ \bibinfo {author} {\bibfnamefont {M.~B.}\ \bibnamefont
  {Plenio}},\ }\href@noop {} {\bibfield  {journal} {\bibinfo  {journal} {Phys.
  Rev. Lett.}\ }\textbf {\bibinfo {volume} {117}},\ \bibinfo {pages} {123602}
  (\bibinfo {year} {2016})}\BibitemShut {NoStop}%
\bibitem [{\citenamefont {Peng}\ \emph {et~al.}(2019)\citenamefont {Peng},
  \citenamefont {Rico}, \citenamefont {Zhong}, \citenamefont {Solano},\ and\
  \citenamefont {Egusquiza}}]{Peng_2019}%
  \BibitemOpen
  \bibfield  {author} {\bibinfo {author} {\bibfnamefont {J.}~\bibnamefont
  {Peng}}, \bibinfo {author} {\bibfnamefont {E.}~\bibnamefont {Rico}}, \bibinfo
  {author} {\bibfnamefont {J.}~\bibnamefont {Zhong}}, \bibinfo {author}
  {\bibfnamefont {E.}~\bibnamefont {Solano}},\ and\ \bibinfo {author}
  {\bibfnamefont {I.~L.}\ \bibnamefont {Egusquiza}},\ }\href@noop {} {\bibfield
   {journal} {\bibinfo  {journal} {Phys. Rev. A}\ }\textbf {\bibinfo {volume}
  {100}},\ \bibinfo {pages} {063820} (\bibinfo {year} {2019})}\BibitemShut
  {NoStop}%
\bibitem [{\citenamefont {Ding}\ \emph {et~al.}(2018)\citenamefont {Ding},
  \citenamefont {Ma}, \citenamefont {Zhang},\ and\ \citenamefont
  {Chan}}]{Ding_2018}%
  \BibitemOpen
  \bibfield  {author} {\bibinfo {author} {\bibfnamefont {K.}~\bibnamefont
  {Ding}}, \bibinfo {author} {\bibfnamefont {G.}~\bibnamefont {Ma}}, \bibinfo
  {author} {\bibfnamefont {Z.~Q.}\ \bibnamefont {Zhang}},\ and\ \bibinfo
  {author} {\bibfnamefont {C.~T.}\ \bibnamefont {Chan}},\ }\href
  {https://doi.org/10.1103/PhysRevLett.121.085702} {\bibfield  {journal}
  {\bibinfo  {journal} {Phys. Rev. Lett.}\ }\textbf {\bibinfo {volume} {121}},\
  \bibinfo {pages} {085702} (\bibinfo {year} {2018})}\BibitemShut {NoStop}%
\bibitem [{\citenamefont {Riwar}(2019)}]{Riwar_2019b}%
  \BibitemOpen
  \bibfield  {author} {\bibinfo {author} {\bibfnamefont {R.-P.}\ \bibnamefont
  {Riwar}},\ }\href
  {https://doi.org/https://doi.org/10.1103/PhysRevB.100.245416} {\bibfield
  {journal} {\bibinfo  {journal} {Phys. Rev. B}\ }\textbf {\bibinfo {volume}
  {100}},\ \bibinfo {pages} {245416} (\bibinfo {year} {2019})}\BibitemShut
  {NoStop}%
\bibitem [{\citenamefont {\"Ozt\"urk}\ \emph {et~al.}(2021)\citenamefont
  {\"Ozt\"urk}, \citenamefont {Lappe}, \citenamefont {Hellmann}, \citenamefont
  {Schmitt}, \citenamefont {Klaers}, \citenamefont {Vewinger}, \citenamefont
  {Kroha},\ and\ \citenamefont {Weitz}}]{Fahri_2021}%
  \BibitemOpen
  \bibfield  {author} {\bibinfo {author} {\bibfnamefont {F.~E.}\ \bibnamefont
  {\"Ozt\"urk}}, \bibinfo {author} {\bibfnamefont {T.}~\bibnamefont {Lappe}},
  \bibinfo {author} {\bibfnamefont {G.}~\bibnamefont {Hellmann}}, \bibinfo
  {author} {\bibfnamefont {J.}~\bibnamefont {Schmitt}}, \bibinfo {author}
  {\bibfnamefont {J.}~\bibnamefont {Klaers}}, \bibinfo {author} {\bibfnamefont
  {F.}~\bibnamefont {Vewinger}}, \bibinfo {author} {\bibfnamefont
  {J.}~\bibnamefont {Kroha}},\ and\ \bibinfo {author} {\bibfnamefont
  {M.}~\bibnamefont {Weitz}},\ }\href {https://doi.org/10.1126/science.abe9869}
  {\bibfield  {journal} {\bibinfo  {journal} {Science}\ }\textbf {\bibinfo
  {volume} {372}},\ \bibinfo {pages} {88} (\bibinfo {year} {2021})},\ \Eprint
  {https://arxiv.org/abs/https://www.science.org/doi/pdf/10.1126/science.abe9869}
  {https://www.science.org/doi/pdf/10.1126/science.abe9869} \BibitemShut
  {NoStop}%
\bibitem [{\citenamefont {Liao}\ \emph {et~al.}(2021)\citenamefont {Liao},
  \citenamefont {Leblanc}, \citenamefont {Ren}, \citenamefont {Li},
  \citenamefont {Li}, \citenamefont {Solnyshkov}, \citenamefont {Malpuech},
  \citenamefont {Yao},\ and\ \citenamefont {Fu}}]{Liao_2021}%
  \BibitemOpen
  \bibfield  {author} {\bibinfo {author} {\bibfnamefont {Q.}~\bibnamefont
  {Liao}}, \bibinfo {author} {\bibfnamefont {C.}~\bibnamefont {Leblanc}},
  \bibinfo {author} {\bibfnamefont {J.}~\bibnamefont {Ren}}, \bibinfo {author}
  {\bibfnamefont {F.}~\bibnamefont {Li}}, \bibinfo {author} {\bibfnamefont
  {Y.}~\bibnamefont {Li}}, \bibinfo {author} {\bibfnamefont {D.}~\bibnamefont
  {Solnyshkov}}, \bibinfo {author} {\bibfnamefont {G.}~\bibnamefont
  {Malpuech}}, \bibinfo {author} {\bibfnamefont {J.}~\bibnamefont {Yao}},\ and\
  \bibinfo {author} {\bibfnamefont {H.}~\bibnamefont {Fu}},\ }\href
  {https://doi.org/10.1103/PhysRevLett.127.107402} {\bibfield  {journal}
  {\bibinfo  {journal} {Phys. Rev. Lett.}\ }\textbf {\bibinfo {volume} {127}},\
  \bibinfo {pages} {107402} (\bibinfo {year} {2021})}\BibitemShut {NoStop}%
\bibitem [{\citenamefont {Hlushchenko}\ \emph {et~al.}(2021)\citenamefont
  {Hlushchenko}, \citenamefont {Novitsky}, \citenamefont {Shcherbinin},\ and\
  \citenamefont {Tuz}}]{Hlushchenko_2021}%
  \BibitemOpen
  \bibfield  {author} {\bibinfo {author} {\bibfnamefont {A.~V.}\ \bibnamefont
  {Hlushchenko}}, \bibinfo {author} {\bibfnamefont {D.~V.}\ \bibnamefont
  {Novitsky}}, \bibinfo {author} {\bibfnamefont {V.~I.}\ \bibnamefont
  {Shcherbinin}},\ and\ \bibinfo {author} {\bibfnamefont {V.~R.}\ \bibnamefont
  {Tuz}},\ }\href {https://doi.org/10.1088/2040-8986/ac31d4} {\bibfield
  {journal} {\bibinfo  {journal} {Journal of Optics}\ }\textbf {\bibinfo
  {volume} {23}},\ \bibinfo {pages} {125002} (\bibinfo {year}
  {2021})}\BibitemShut {NoStop}%
\bibitem [{\citenamefont {Xia}\ \emph {et~al.}(2021)\citenamefont {Xia},
  \citenamefont {Danieli}, \citenamefont {Zhang}, \citenamefont {Zhao},
  \citenamefont {Lu}, \citenamefont {Tang}, \citenamefont {Li}, \citenamefont
  {Song},\ and\ \citenamefont {Chen}}]{Shiqiang_2021}%
  \BibitemOpen
  \bibfield  {author} {\bibinfo {author} {\bibfnamefont {S.}~\bibnamefont
  {Xia}}, \bibinfo {author} {\bibfnamefont {C.}~\bibnamefont {Danieli}},
  \bibinfo {author} {\bibfnamefont {Y.}~\bibnamefont {Zhang}}, \bibinfo
  {author} {\bibfnamefont {X.}~\bibnamefont {Zhao}}, \bibinfo {author}
  {\bibfnamefont {H.}~\bibnamefont {Lu}}, \bibinfo {author} {\bibfnamefont
  {L.}~\bibnamefont {Tang}}, \bibinfo {author} {\bibfnamefont {D.}~\bibnamefont
  {Li}}, \bibinfo {author} {\bibfnamefont {D.}~\bibnamefont {Song}},\ and\
  \bibinfo {author} {\bibfnamefont {Z.}~\bibnamefont {Chen}},\ }\href
  {https://doi.org/10.1063/5.0069633} {\bibfield  {journal} {\bibinfo
  {journal} {APL Photonics}\ }\textbf {\bibinfo {volume} {6}},\ \bibinfo
  {pages} {126106} (\bibinfo {year} {2021})},\ \Eprint
  {https://arxiv.org/abs/https://pubs.aip.org/aip/app/article-pdf/doi/10.1063/5.0069633/14055975/126106\_1\_online.pdf}
  {https://pubs.aip.org/aip/app/article-pdf/doi/10.1063/5.0069633/14055975/126106\_1\_online.pdf}
  \BibitemShut {NoStop}%
\bibitem [{\citenamefont {Ding}\ \emph {et~al.}(2022)\citenamefont {Ding},
  \citenamefont {Fang},\ and\ \citenamefont {Ma}}]{Ding_2022}%
  \BibitemOpen
  \bibfield  {author} {\bibinfo {author} {\bibfnamefont {K.}~\bibnamefont
  {Ding}}, \bibinfo {author} {\bibfnamefont {C.}~\bibnamefont {Fang}},\ and\
  \bibinfo {author} {\bibfnamefont {G.}~\bibnamefont {Ma}},\ }\href@noop {}
  {\bibfield  {journal} {\bibinfo  {journal} {Nature Reviews Physics}\ }\textbf
  {\bibinfo {volume} {4}},\ \bibinfo {pages} {745} (\bibinfo {year}
  {2022})}\BibitemShut {NoStop}%
\bibitem [{\citenamefont {Javed}\ \emph {et~al.}(2023)\citenamefont {Javed},
  \citenamefont {Schwibbert},\ and\ \citenamefont {Riwar}}]{Javed_2023}%
  \BibitemOpen
  \bibfield  {author} {\bibinfo {author} {\bibfnamefont {M.~A.}\ \bibnamefont
  {Javed}}, \bibinfo {author} {\bibfnamefont {J.}~\bibnamefont {Schwibbert}},\
  and\ \bibinfo {author} {\bibfnamefont {R.-P.}\ \bibnamefont {Riwar}},\ }\href
  {https://doi.org/10.1103/PhysRevB.107.035408} {\bibfield  {journal} {\bibinfo
   {journal} {Phys. Rev. B}\ }\textbf {\bibinfo {volume} {107}},\ \bibinfo
  {pages} {035408} (\bibinfo {year} {2023})}\BibitemShut {NoStop}%
\bibitem [{\citenamefont {Kunst}\ \emph {et~al.}(2018)\citenamefont {Kunst},
  \citenamefont {Edvardsson}, \citenamefont {Budich},\ and\ \citenamefont
  {Bergholtz}}]{Kunst:2018aa}%
  \BibitemOpen
  \bibfield  {author} {\bibinfo {author} {\bibfnamefont {F.~K.}\ \bibnamefont
  {Kunst}}, \bibinfo {author} {\bibfnamefont {E.}~\bibnamefont {Edvardsson}},
  \bibinfo {author} {\bibfnamefont {J.~C.}\ \bibnamefont {Budich}},\ and\
  \bibinfo {author} {\bibfnamefont {E.~J.}\ \bibnamefont {Bergholtz}},\ }\href
  {https://doi.org/10.1103/PhysRevLett.121.026808} {\bibfield  {journal}
  {\bibinfo  {journal} {Physical Review Letters}\ }\textbf {\bibinfo {volume}
  {121}},\ \bibinfo {pages} {026808} (\bibinfo {year} {2018})}\BibitemShut
  {NoStop}%
\bibitem [{\citenamefont {Kawabata}\ \emph {et~al.}(2019)\citenamefont
  {Kawabata}, \citenamefont {Shiozaki}, \citenamefont {Ueda},\ and\
  \citenamefont {Sato}}]{Kawabata:2019aa}%
  \BibitemOpen
  \bibfield  {author} {\bibinfo {author} {\bibfnamefont {K.}~\bibnamefont
  {Kawabata}}, \bibinfo {author} {\bibfnamefont {K.}~\bibnamefont {Shiozaki}},
  \bibinfo {author} {\bibfnamefont {M.}~\bibnamefont {Ueda}},\ and\ \bibinfo
  {author} {\bibfnamefont {M.}~\bibnamefont {Sato}},\ }\href
  {https://doi.org/10.1103/PhysRevX.9.041015} {\bibfield  {journal} {\bibinfo
  {journal} {Physical Review X}\ }\textbf {\bibinfo {volume} {9}},\ \bibinfo
  {pages} {041015} (\bibinfo {year} {2019})}\BibitemShut {NoStop}%
\bibitem [{\citenamefont {Heiss}(2004)}]{Heiss_2004}%
  \BibitemOpen
  \bibfield  {author} {\bibinfo {author} {\bibfnamefont {W.~D.}\ \bibnamefont
  {Heiss}},\ }\href {https://doi.org/10.1088/0305-4470/37/6/034} {\bibfield
  {journal} {\bibinfo  {journal} {Journal of Physics A: Mathematical and
  General}\ }\textbf {\bibinfo {volume} {37}},\ \bibinfo {pages} {2455}
  (\bibinfo {year} {2004})}\BibitemShut {NoStop}%
\bibitem [{\citenamefont {Heiss}(2012)}]{Heiss_2012}%
  \BibitemOpen
  \bibfield  {author} {\bibinfo {author} {\bibfnamefont {W.~D.}\ \bibnamefont
  {Heiss}},\ }\href {https://doi.org/10.1088/1751-8113/45/44/444016} {\bibfield
   {journal} {\bibinfo  {journal} {Journal of Physics A: Mathematical and
  Theoretical}\ }\textbf {\bibinfo {volume} {45}},\ \bibinfo {pages} {444016}
  (\bibinfo {year} {2012})}\BibitemShut {NoStop}%
\bibitem [{\citenamefont {Wu}\ \emph {et~al.}(2019)\citenamefont {Wu},
  \citenamefont {Jin},\ and\ \citenamefont {Song}}]{Wu_2019}%
  \BibitemOpen
  \bibfield  {author} {\bibinfo {author} {\bibfnamefont {H.~C.}\ \bibnamefont
  {Wu}}, \bibinfo {author} {\bibfnamefont {L.}~\bibnamefont {Jin}},\ and\
  \bibinfo {author} {\bibfnamefont {Z.}~\bibnamefont {Song}},\ }\href
  {https://doi.org/10.1103/PhysRevB.100.155117} {\bibfield  {journal} {\bibinfo
   {journal} {Phys. Rev. B}\ }\textbf {\bibinfo {volume} {100}},\ \bibinfo
  {pages} {155117} (\bibinfo {year} {2019})}\BibitemShut {NoStop}%
\bibitem [{\citenamefont {Mandal}\ and\ \citenamefont
  {Bergholtz}(2021)}]{Mandal_2021}%
  \BibitemOpen
  \bibfield  {author} {\bibinfo {author} {\bibfnamefont {I.}~\bibnamefont
  {Mandal}}\ and\ \bibinfo {author} {\bibfnamefont {E.~J.}\ \bibnamefont
  {Bergholtz}},\ }\href {https://doi.org/10.1103/PhysRevLett.127.186601}
  {\bibfield  {journal} {\bibinfo  {journal} {Phys. Rev. Lett.}\ }\textbf
  {\bibinfo {volume} {127}},\ \bibinfo {pages} {186601} (\bibinfo {year}
  {2021})}\BibitemShut {NoStop}%
\bibitem [{\citenamefont {Bergholtz}\ \emph {et~al.}(2021)\citenamefont
  {Bergholtz}, \citenamefont {Budich},\ and\ \citenamefont
  {Kunst}}]{Bergholtz_2021}%
  \BibitemOpen
  \bibfield  {author} {\bibinfo {author} {\bibfnamefont {E.~J.}\ \bibnamefont
  {Bergholtz}}, \bibinfo {author} {\bibfnamefont {J.~C.}\ \bibnamefont
  {Budich}},\ and\ \bibinfo {author} {\bibfnamefont {F.~K.}\ \bibnamefont
  {Kunst}},\ }\href {https://doi.org/10.1103/RevModPhys.93.015005} {\bibfield
  {journal} {\bibinfo  {journal} {Rev. Mod. Phys.}\ }\textbf {\bibinfo {volume}
  {93}},\ \bibinfo {pages} {015005} (\bibinfo {year} {2021})}\BibitemShut
  {NoStop}%
\bibitem [{\citenamefont {Avila}\ \emph {et~al.}(2019)\citenamefont {Avila},
  \citenamefont {Pe{\~{n}}aranda}, \citenamefont {Prada}, \citenamefont
  {San-Jose},\ and\ \citenamefont {Aguado}}]{Avila_2019}%
  \BibitemOpen
  \bibfield  {author} {\bibinfo {author} {\bibfnamefont {J.}~\bibnamefont
  {Avila}}, \bibinfo {author} {\bibfnamefont {F.}~\bibnamefont
  {Pe{\~{n}}aranda}}, \bibinfo {author} {\bibfnamefont {E.}~\bibnamefont
  {Prada}}, \bibinfo {author} {\bibfnamefont {P.}~\bibnamefont {San-Jose}},\
  and\ \bibinfo {author} {\bibfnamefont {R.}~\bibnamefont {Aguado}},\
  }\bibfield  {journal} {\bibinfo  {journal} {Communications Physics}\ }\textbf
  {\bibinfo {volume} {2}},\ \href {https://doi.org/10.1038/s42005-019-0231-8}
  {10.1038/s42005-019-0231-8} (\bibinfo {year} {2019})\BibitemShut {NoStop}%
\bibitem [{\citenamefont {San-Jose}\ \emph {et~al.}(2016)\citenamefont
  {San-Jose}, \citenamefont {Cayao}, \citenamefont {Prada},\ and\ \citenamefont
  {Aguado}}]{San-Jose_2016}%
  \BibitemOpen
  \bibfield  {author} {\bibinfo {author} {\bibfnamefont {P.}~\bibnamefont
  {San-Jose}}, \bibinfo {author} {\bibfnamefont {J.}~\bibnamefont {Cayao}},
  \bibinfo {author} {\bibfnamefont {E.}~\bibnamefont {Prada}},\ and\ \bibinfo
  {author} {\bibfnamefont {R.}~\bibnamefont {Aguado}},\ }\bibfield  {journal}
  {\bibinfo  {journal} {Scientific Reports}\ }\textbf {\bibinfo {volume} {6}},\
  \href {https://doi.org/10.1038/srep21427} {10.1038/srep21427} (\bibinfo
  {year} {2016})\BibitemShut {NoStop}%
\bibitem [{\citenamefont {Khandelwal}\ \emph {et~al.}(2021)\citenamefont
  {Khandelwal}, \citenamefont {Brunner},\ and\ \citenamefont
  {Haack}}]{Khandelwal_2021}%
  \BibitemOpen
  \bibfield  {author} {\bibinfo {author} {\bibfnamefont {S.}~\bibnamefont
  {Khandelwal}}, \bibinfo {author} {\bibfnamefont {N.}~\bibnamefont
  {Brunner}},\ and\ \bibinfo {author} {\bibfnamefont {G.}~\bibnamefont
  {Haack}},\ }\href {https://doi.org/10.1103/PRXQuantum.2.040346} {\bibfield
  {journal} {\bibinfo  {journal} {PRX Quantum}\ }\textbf {\bibinfo {volume}
  {2}},\ \bibinfo {pages} {040346} (\bibinfo {year} {2021})}\BibitemShut
  {NoStop}%
\bibitem [{\citenamefont {Khandelwal}\ \emph {et~al.}(2024)\citenamefont
  {Khandelwal}, \citenamefont {Chen}, \citenamefont {Murch},\ and\
  \citenamefont {Haack}}]{Khandelwal_2024}%
  \BibitemOpen
  \bibfield  {author} {\bibinfo {author} {\bibfnamefont {S.}~\bibnamefont
  {Khandelwal}}, \bibinfo {author} {\bibfnamefont {W.}~\bibnamefont {Chen}},
  \bibinfo {author} {\bibfnamefont {K.~W.}\ \bibnamefont {Murch}},\ and\
  \bibinfo {author} {\bibfnamefont {G.}~\bibnamefont {Haack}},\ }\href
  {https://doi.org/10.1103/PhysRevLett.133.070403} {\bibfield  {journal}
  {\bibinfo  {journal} {Phys. Rev. Lett.}\ }\textbf {\bibinfo {volume} {133}},\
  \bibinfo {pages} {070403} (\bibinfo {year} {2024})}\BibitemShut {NoStop}%
\bibitem [{\citenamefont {Schmid}(1983)}]{Schmid_1983}%
  \BibitemOpen
  \bibfield  {author} {\bibinfo {author} {\bibfnamefont {A.}~\bibnamefont
  {Schmid}},\ }\href {https://doi.org/10.1103/PhysRevLett.51.1506} {\bibfield
  {journal} {\bibinfo  {journal} {Phys. Rev. Lett.}\ }\textbf {\bibinfo
  {volume} {51}},\ \bibinfo {pages} {1506} (\bibinfo {year}
  {1983})}\BibitemShut {NoStop}%
\bibitem [{\citenamefont {Bulgadaev}(1984)}]{Bulgadaev_1984}%
  \BibitemOpen
  \bibfield  {author} {\bibinfo {author} {\bibfnamefont {S.~A.}\ \bibnamefont
  {Bulgadaev}},\ }\href@noop {} {\bibfield  {journal} {\bibinfo  {journal}
  {JETP Letters}\ }\textbf {\bibinfo {volume} {39}},\ \bibinfo {pages} {315}
  (\bibinfo {year} {1984})}\BibitemShut {NoStop}%
\bibitem [{\citenamefont {Leggett}\ \emph {et~al.}(1987)\citenamefont
  {Leggett}, \citenamefont {Chakravarty}, \citenamefont {Dorsey}, \citenamefont
  {Fisher}, \citenamefont {Garg},\ and\ \citenamefont
  {Zwerger}}]{Leggett_1987}%
  \BibitemOpen
  \bibfield  {author} {\bibinfo {author} {\bibfnamefont {A.~J.}\ \bibnamefont
  {Leggett}}, \bibinfo {author} {\bibfnamefont {S.}~\bibnamefont
  {Chakravarty}}, \bibinfo {author} {\bibfnamefont {A.~T.}\ \bibnamefont
  {Dorsey}}, \bibinfo {author} {\bibfnamefont {M.~P.~A.}\ \bibnamefont
  {Fisher}}, \bibinfo {author} {\bibfnamefont {A.}~\bibnamefont {Garg}},\ and\
  \bibinfo {author} {\bibfnamefont {W.}~\bibnamefont {Zwerger}},\ }\href
  {https://doi.org/10.1103/RevModPhys.59.1} {\bibfield  {journal} {\bibinfo
  {journal} {Rev. Mod. Phys.}\ }\textbf {\bibinfo {volume} {59}},\ \bibinfo
  {pages} {1} (\bibinfo {year} {1987})}\BibitemShut {NoStop}%
\bibitem [{\citenamefont {Murani}\ \emph {et~al.}(2020)\citenamefont {Murani},
  \citenamefont {Bourlet}, \citenamefont {le~Sueur}, \citenamefont {Portier},
  \citenamefont {Altimiras}, \citenamefont {Esteve}, \citenamefont {Grabert},
  \citenamefont {Stockburger}, \citenamefont {Ankerhold},\ and\ \citenamefont
  {Joyez}}]{Murani_2020}%
  \BibitemOpen
  \bibfield  {author} {\bibinfo {author} {\bibfnamefont {A.}~\bibnamefont
  {Murani}}, \bibinfo {author} {\bibfnamefont {N.}~\bibnamefont {Bourlet}},
  \bibinfo {author} {\bibfnamefont {H.}~\bibnamefont {le~Sueur}}, \bibinfo
  {author} {\bibfnamefont {F.}~\bibnamefont {Portier}}, \bibinfo {author}
  {\bibfnamefont {C.}~\bibnamefont {Altimiras}}, \bibinfo {author}
  {\bibfnamefont {D.}~\bibnamefont {Esteve}}, \bibinfo {author} {\bibfnamefont
  {H.}~\bibnamefont {Grabert}}, \bibinfo {author} {\bibfnamefont
  {J.}~\bibnamefont {Stockburger}}, \bibinfo {author} {\bibfnamefont
  {J.}~\bibnamefont {Ankerhold}},\ and\ \bibinfo {author} {\bibfnamefont
  {P.}~\bibnamefont {Joyez}},\ }\href
  {https://doi.org/10.1103/PhysRevX.10.021003} {\bibfield  {journal} {\bibinfo
  {journal} {Phys. Rev. X}\ }\textbf {\bibinfo {volume} {10}},\ \bibinfo
  {pages} {021003} (\bibinfo {year} {2020})}\BibitemShut {NoStop}%
\bibitem [{\citenamefont {Hakonen}\ and\ \citenamefont
  {Sonin}(2021)}]{Hakonen_2021}%
  \BibitemOpen
  \bibfield  {author} {\bibinfo {author} {\bibfnamefont {P.~J.}\ \bibnamefont
  {Hakonen}}\ and\ \bibinfo {author} {\bibfnamefont {E.~B.}\ \bibnamefont
  {Sonin}},\ }\href
  {https://doi.org/https://doi.org/10.1103/PhysRevX.11.018001} {\bibfield
  {journal} {\bibinfo  {journal} {Phys. Rev. X}\ }\textbf {\bibinfo {volume}
  {11}},\ \bibinfo {pages} {018001} (\bibinfo {year} {2021})}\BibitemShut
  {NoStop}%
\bibitem [{\citenamefont {Murani}\ \emph {et~al.}(2021)\citenamefont {Murani},
  \citenamefont {Bourlet}, \citenamefont {le~Sueur}, \citenamefont {Portier},
  \citenamefont {Altimiras}, \citenamefont {Esteve}, \citenamefont {Grabert},
  \citenamefont {Stockburger}, \citenamefont {Ankerhold},\ and\ \citenamefont
  {Joyez}}]{Murani_2021}%
  \BibitemOpen
  \bibfield  {author} {\bibinfo {author} {\bibfnamefont {A.}~\bibnamefont
  {Murani}}, \bibinfo {author} {\bibfnamefont {N.}~\bibnamefont {Bourlet}},
  \bibinfo {author} {\bibfnamefont {H.}~\bibnamefont {le~Sueur}}, \bibinfo
  {author} {\bibfnamefont {F.}~\bibnamefont {Portier}}, \bibinfo {author}
  {\bibfnamefont {C.}~\bibnamefont {Altimiras}}, \bibinfo {author}
  {\bibfnamefont {D.}~\bibnamefont {Esteve}}, \bibinfo {author} {\bibfnamefont
  {H.}~\bibnamefont {Grabert}}, \bibinfo {author} {\bibfnamefont
  {J.}~\bibnamefont {Stockburger}}, \bibinfo {author} {\bibfnamefont
  {J.}~\bibnamefont {Ankerhold}},\ and\ \bibinfo {author} {\bibfnamefont
  {P.}~\bibnamefont {Joyez}},\ }\href
  {https://doi.org/10.1103/PhysRevX.11.018002} {\bibfield  {journal} {\bibinfo
  {journal} {Phys. Rev. X}\ }\textbf {\bibinfo {volume} {11}},\ \bibinfo
  {pages} {018002} (\bibinfo {year} {2021})}\BibitemShut {NoStop}%
\bibitem [{\citenamefont {Kuzmin}\ \emph {et~al.}()\citenamefont {Kuzmin},
  \citenamefont {Mehta}, \citenamefont {Grabon}, \citenamefont {Mencia},
  \citenamefont {Burshtein}, \citenamefont {Goldstein},\ and\ \citenamefont
  {Manucharyan}}]{kuzmin2023observation}%
  \BibitemOpen
  \bibfield  {author} {\bibinfo {author} {\bibfnamefont {R.}~\bibnamefont
  {Kuzmin}}, \bibinfo {author} {\bibfnamefont {N.}~\bibnamefont {Mehta}},
  \bibinfo {author} {\bibfnamefont {N.}~\bibnamefont {Grabon}}, \bibinfo
  {author} {\bibfnamefont {R.~A.}\ \bibnamefont {Mencia}}, \bibinfo {author}
  {\bibfnamefont {A.}~\bibnamefont {Burshtein}}, \bibinfo {author}
  {\bibfnamefont {M.}~\bibnamefont {Goldstein}},\ and\ \bibinfo {author}
  {\bibfnamefont {V.~E.}\ \bibnamefont {Manucharyan}},\ }\href@noop {}
  {}\Eprint {https://arxiv.org/abs/2304.05806} {arXiv:2304.05806} \BibitemShut
  {NoStop}%
\bibitem [{\citenamefont {Houzet}\ \emph {et~al.}(2024)\citenamefont {Houzet},
  \citenamefont {Yamamoto},\ and\ \citenamefont
  {Glazman}}]{houzet2024microwave}%
  \BibitemOpen
  \bibfield  {author} {\bibinfo {author} {\bibfnamefont {M.}~\bibnamefont
  {Houzet}}, \bibinfo {author} {\bibfnamefont {T.}~\bibnamefont {Yamamoto}},\
  and\ \bibinfo {author} {\bibfnamefont {L.~I.}\ \bibnamefont {Glazman}},\
  }\href {https://doi.org/10.1103/PhysRevB.109.155431} {\bibfield  {journal}
  {\bibinfo  {journal} {Phys. Rev. B}\ }\textbf {\bibinfo {volume} {109}},\
  \bibinfo {pages} {155431} (\bibinfo {year} {2024})}\BibitemShut {NoStop}%
\bibitem [{\citenamefont {Burshtein}\ and\ \citenamefont
  {Goldstein}(2024)}]{burshtein2024inelastic}%
  \BibitemOpen
  \bibfield  {author} {\bibinfo {author} {\bibfnamefont {A.}~\bibnamefont
  {Burshtein}}\ and\ \bibinfo {author} {\bibfnamefont {M.}~\bibnamefont
  {Goldstein}},\ }\href {https://doi.org/10.1103/PRXQuantum.5.020323}
  {\bibfield  {journal} {\bibinfo  {journal} {PRX Quantum}\ }\textbf {\bibinfo
  {volume} {5}},\ \bibinfo {pages} {020323} (\bibinfo {year}
  {2024})}\BibitemShut {NoStop}%
\bibitem [{\citenamefont {Masuki}\ \emph {et~al.}(2022)\citenamefont {Masuki},
  \citenamefont {Sudo}, \citenamefont {Oshikawa},\ and\ \citenamefont
  {Ashida}}]{Masuki_2022}%
  \BibitemOpen
  \bibfield  {author} {\bibinfo {author} {\bibfnamefont {K.}~\bibnamefont
  {Masuki}}, \bibinfo {author} {\bibfnamefont {H.}~\bibnamefont {Sudo}},
  \bibinfo {author} {\bibfnamefont {M.}~\bibnamefont {Oshikawa}},\ and\
  \bibinfo {author} {\bibfnamefont {Y.}~\bibnamefont {Ashida}},\ }\href
  {https://doi.org/10.1103/PhysRevLett.129.087001} {\bibfield  {journal}
  {\bibinfo  {journal} {Phys. Rev. Lett.}\ }\textbf {\bibinfo {volume} {129}},\
  \bibinfo {pages} {087001} (\bibinfo {year} {2022})}\BibitemShut {NoStop}%
\bibitem [{\citenamefont {S\'epulcre}\ \emph {et~al.}(2023)\citenamefont
  {S\'epulcre}, \citenamefont {Florens},\ and\ \citenamefont
  {Snyman}}]{Sepulcre_2023}%
  \BibitemOpen
  \bibfield  {author} {\bibinfo {author} {\bibfnamefont {T.}~\bibnamefont
  {S\'epulcre}}, \bibinfo {author} {\bibfnamefont {S.}~\bibnamefont
  {Florens}},\ and\ \bibinfo {author} {\bibfnamefont {I.}~\bibnamefont
  {Snyman}},\ }\href {https://doi.org/10.1103/PhysRevLett.131.199701}
  {\bibfield  {journal} {\bibinfo  {journal} {Phys. Rev. Lett.}\ }\textbf
  {\bibinfo {volume} {131}},\ \bibinfo {pages} {199701} (\bibinfo {year}
  {2023})}\BibitemShut {NoStop}%
\bibitem [{\citenamefont {Masuki}\ \emph {et~al.}(2023)\citenamefont {Masuki},
  \citenamefont {Sudo}, \citenamefont {Oshikawa},\ and\ \citenamefont
  {Ashida}}]{Masuki_2023}%
  \BibitemOpen
  \bibfield  {author} {\bibinfo {author} {\bibfnamefont {K.}~\bibnamefont
  {Masuki}}, \bibinfo {author} {\bibfnamefont {H.}~\bibnamefont {Sudo}},
  \bibinfo {author} {\bibfnamefont {M.}~\bibnamefont {Oshikawa}},\ and\
  \bibinfo {author} {\bibfnamefont {Y.}~\bibnamefont {Ashida}},\ }\href
  {https://doi.org/10.1103/PhysRevLett.131.199702} {\bibfield  {journal}
  {\bibinfo  {journal} {Phys. Rev. Lett.}\ }\textbf {\bibinfo {volume} {131}},\
  \bibinfo {pages} {199702} (\bibinfo {year} {2023})}\BibitemShut {NoStop}%
\bibitem [{\citenamefont {Giacomelli}\ and\ \citenamefont
  {Ciuti}(2024)}]{giacomelli2024emergent}%
  \BibitemOpen
  \bibfield  {author} {\bibinfo {author} {\bibfnamefont {L.}~\bibnamefont
  {Giacomelli}}\ and\ \bibinfo {author} {\bibfnamefont {C.}~\bibnamefont
  {Ciuti}},\ }\href {https://doi.org/10.1038/s41467-024-48558-w} {\bibfield
  {journal} {\bibinfo  {journal} {Nature Communications}\ }\textbf {\bibinfo
  {volume} {15}},\ \bibinfo {pages} {5455} (\bibinfo {year}
  {2024})}\BibitemShut {NoStop}%
\bibitem [{\citenamefont {Kashuba}\ and\ \citenamefont
  {Riwar}(2024)}]{kashuba2024phasetransitions}%
  \BibitemOpen
  \bibfield  {author} {\bibinfo {author} {\bibfnamefont {O.}~\bibnamefont
  {Kashuba}}\ and\ \bibinfo {author} {\bibfnamefont {R.-P.}\ \bibnamefont
  {Riwar}},\ }\href {https://doi.org/10.1103/PhysRevB.110.184505} {\bibfield
  {journal} {\bibinfo  {journal} {Phys. Rev. B}\ }\textbf {\bibinfo {volume}
  {110}},\ \bibinfo {pages} {184505} (\bibinfo {year} {2024})}\BibitemShut
  {NoStop}%
\bibitem [{\citenamefont {Placke}\ \emph {et~al.}(2018)\citenamefont {Placke},
  \citenamefont {Pluecker}, \citenamefont {Splettstoesser},\ and\ \citenamefont
  {Wegewijs}}]{Placke_2018}%
  \BibitemOpen
  \bibfield  {author} {\bibinfo {author} {\bibfnamefont {B.~A.}\ \bibnamefont
  {Placke}}, \bibinfo {author} {\bibfnamefont {T.}~\bibnamefont {Pluecker}},
  \bibinfo {author} {\bibfnamefont {J.}~\bibnamefont {Splettstoesser}},\ and\
  \bibinfo {author} {\bibfnamefont {M.~R.}\ \bibnamefont {Wegewijs}},\ }\href
  {https://doi.org/https://doi.org/10.1103/PhysRevB.98.085307} {\bibfield
  {journal} {\bibinfo  {journal} {Phys. Rev. B}\ }\textbf {\bibinfo {volume}
  {98}},\ \bibinfo {pages} {085307} (\bibinfo {year} {2018})}\BibitemShut
  {NoStop}%
\bibitem [{Note4()}]{Note4}%
  \BibitemOpen
  \bibinfo {note} {The meaning and usefulness of such an incomplete Legendre
  transformation, especially in the context of the PI treatment, will become
  clear throughout this work.}\BibitemShut {Stop}%
\bibitem [{Note5()}]{Note5}%
  \BibitemOpen
  \bibinfo {note} {By standard Legendre transformation, we mean $N=\partial
  _{\protect \dot {\phi }}L$ and $H=N\protect \dot {\phi }-L$ when going from
  $L$ to $H$, and vice versa from $H$ to $L$.}\BibitemShut {Stop}%
\bibitem [{Note6()}]{Note6}%
  \BibitemOpen
  \bibinfo {note} {For simplicity, this work will mostly focus on linear
  inductors, where the problem of nonconvex $T(N)$ could in principle be
  avoided by performing a Legendre transformation on $V(\phi )$ instead of
  $T(N)$, similar in spirit to Ref.~\cite {Ulrich_2016}. But our goal is to
  provide a theory that works independent of whether the inductive shunt is
  linear or nonlinear.}\BibitemShut {Stop}%
\bibitem [{Note7()}]{Note7}%
  \BibitemOpen
  \bibinfo {note} {Note that the regular $\sim \protect \qopname \relax
  o{cos}(\phi )$ energy contribution from a standard Josephson tunneling
  junction follows from the exact same low-energy approximation of the full BCS
  description of the junction, within the Lagrangian. The reason for selecting
  the highest, and not the lowest eigenvalue becomes clear when considering the
  system in a thermal Gibbs state (for which we will use imaginary times), see
  further below.}\BibitemShut {Stop}%
\bibitem [{Sup()}]{Supplemental_Material}%
  \BibitemOpen
  \href@noop {} {}\bibinfo {note} {See Supplemental Material for more
  details.}\BibitemShut {Stop}%
\bibitem [{Note8()}]{Note8}%
  \BibitemOpen
  \bibinfo {note} {Via analytic continuation from imaginary times back to real
  times, we now understand why in Eq.~\protect \textup {\hbox {\mathsurround
  \z@ \protect \normalfont (\ignorespaces \ref {eq_TL_low}\unskip \@@italiccorr
  )}} we had to choose the higher, and not the lower eigenenergy.}\BibitemShut
  {Stop}%
\bibitem [{Note9()}]{Note9}%
  \BibitemOpen
  \bibinfo {note} {Both $\Lambda $ and $\Gamma $ are Hermitean
  operators.}\BibitemShut {Stop}%
\bibitem [{\citenamefont {Rojo}(2003)}]{Rojo_2003}%
  \BibitemOpen
  \bibfield  {author} {\bibinfo {author} {\bibfnamefont {A.~G.}\ \bibnamefont
  {Rojo}},\ }\href {https://doi.org/10.1103/PhysRevA.68.013807} {\bibfield
  {journal} {\bibinfo  {journal} {Phys. Rev. A}\ }\textbf {\bibinfo {volume}
  {68}},\ \bibinfo {pages} {013807} (\bibinfo {year} {2003})}\BibitemShut
  {NoStop}%
\bibitem [{Note10()}]{Note10}%
  \BibitemOpen
  \bibinfo {note} {As an aside: for $J>0$, if $S_z$ remains sufficiently small
  compared to $S$, we can approximate the dynamics as an effective Josephson
  effect, $S_z\sim n$, $S_x\sim \protect \qopname \relax o{cos}(\varphi )$ with
  $[n,\varphi ]=i$. Incidentally, this results in a model that has been
  recently proposed to exist when coupling charge islands to charge qubits,
  resulting in a quasiperiodic non-linear capacitor~\cite
  {Herrig_2023,Herrig_2025}.}\BibitemShut {Stop}%
\end{thebibliography}%

\end{document}


\title{Supplemental Material: Consistent quantum treatments of anharmonic kinetic energies}
\author{C. Koliofoti, M. A. Javed and R.-P. Riwar}
\affiliation{Peter Gr\"unberg Institute, Theoretical Nanoelectronics, Forschungszentrum J\"ulich, D-52425 J\"ulich, Germany}

\begin{abstract}
    This supplemental material provides additional information on the following subjects. In Sec.~\ref{sec_heat_capacity} we detail the calculation of the heat capacitance reported on in the main text, both without and with a resistive shunt. Section~\ref{sec_partial_path_integral} presents the 'partial' path integral approach, wherein the charge and phase operators are turned into classical coordinates, whereas the intrinsic capacitor degrees of freedom remain quantum. In Sec.~\ref{sec_noise_power} we define and compute the path integral noise power spectra for the voltage $\dot{\phi}$ in imaginary time, which is used in the main text to estimate adiabaticity with respect to exceptional points. Section~\ref{sec_basis_transformation} details the imaginary time basis transformation which allows us to deduce a high frequency cutoff in the noise power spectrum. In Sec.~\ref{sec_large_dipole_model} we show how the large pseudo-spin model in the second part of the main text follows from a microscopic model with an ensemble of two-level quantum dipole moments.
\end{abstract}

\maketitle

\section{Calculation of heat capacity}\label{sec_heat_capacity}

\subsection{Zero dissipation}

The partition function can be directly related to the density of states,
which we assume to satisfy a low-energy power law (we rescaled the
energy spectrum such that the ground state is at zero energy)
\begin{equation}
Z =\text{tr}\left[e^{-\beta H}\right]=\int_{0}^{\infty}dE\rho\left(E\right)e^{-\beta E} \sim\int_{0}^{\infty}dEE^{\kappa}e^{-\beta E}=\beta^{-\kappa-1}\Gamma\left(\kappa+1\right)
\end{equation}
where $\Gamma$ is the Gamma function. The heat capacity, defined
as $C_{v}=\beta^{2}\partial_{\beta}^{2}\ln\left(Z\right)$, then straightforwardly
yields
\begin{equation}
C_{v}=1+\kappa.
\end{equation}
We thus need to find the power law of the density of states.

Take a low energy Hamiltonian with kinetic and potential energies
\begin{equation}
H=T\left(N\right)+V\left(\varphi\right).
\end{equation}
We assume (as in the main text) a harmonic potential energy $V\left(\varphi\right)=\phi^{2}/2l$.
As for the kinetic energy, we assume that at low energies, it satisfies
a power law behaviour $T\left(N\right)\sim\left|N\right|^{\alpha}$.
Then, the low-energy density of states in the limit $l\rightarrow\infty$
can be computed by the convolution formula~\cite{Herrig_2025}
\begin{equation}
\rho\left(E\right)\sim\int_{0}^{E}dE^{\prime}\frac{1}{\sqrt{E-E^{\prime}}}\rho_{0}\left(E^{\prime}\right)
\end{equation}
where $\rho_{0}$ is the density of states for the kinetic term only,
assuming equidistant distribution of the eigenvalues $N$,
\begin{equation}
\rho_{0}\sim E^{1-\frac{1}{\alpha}}.
\end{equation}
With the above convolution formula, we get
\begin{equation}
\rho\sim E^{\frac{1}{\alpha}-\frac{1}{2}},
\end{equation}
and, consequently, arrive at
\begin{equation}
C_{v}=\frac{1}{2}+\frac{1}{\alpha}.
\end{equation}
For the kinetic energy of the two-level model in the main text,
\begin{align*}
T\left(N\right) & =\frac{N^{2}}{2c}-\sqrt{\frac{\lambda^{2}N^{2}}{c^{2}}+\gamma^{2}}
\end{align*}
the system undergoes a phase transition at $\gamma_{0}=\lambda^{2}/c$.
Before and after the transition, the power law is $\alpha=2$, yielding
\begin{equation}
C_{v}=1.
\end{equation}
At the transition, the quadratic term vanishes, and we are left with
a low-energy kinetic term that scales with $\alpha=4$, such that
\begin{equation}
C_{v}=\frac{3}{4},
\end{equation}
as stated in the main text.

\subsection{Finite dissipation}

We here compute the expression for the heat capacity in the presence
of an Ohmic resistor. This problem can be expressed in terms of standard
path integral methods, where the partition function in Fourier space
is given as~\cite{Altland_Simons_book}
\begin{equation}\label{eq_supp_path_integral}
Z\sim\lim_{k_{\text{co}}\rightarrow\infty}\int d\phi_{0}e^{-\beta\frac{\phi_{0}^{2}}{2l}}\prod_{k=1}^{k_{\text{co}}}\int d\phi_{kr}\int d\phi_{ki}e^{-\beta\sum_{k=1}^{\frac{K-1}{2}}\left(\frac{c\left(\frac{2\pi k}{\beta}\right)^{2}}{2}+\frac{1}{4\pi\eta}\frac{2\pi k}{\beta}+\frac{1}{2l}\right)\left(\phi_{kr}^{2}+\phi_{ki}^{2}\right)}
\end{equation}
with the Matsubara frequencies $\omega_{k}=2\pi k/\beta$, where for
$k>0$, we take into account the fact that the Fourier components
are complex, $\phi_{k}=\phi_{kr}+i\phi_{ki}$, i.e., the transformation
from imaginary time to frequency space is
\begin{equation}\label{eq_supp_phitau_Matsubara}
\phi_{\tau}=\phi_{0}+\frac{1}{\sqrt{2}}\sum_{k=1}^{k_{\text{co}}}\left[\left(\phi_{kr}+i\phi_{ki}\right)e^{i\frac{2\pi}{\beta}k\tau}+\left(\phi_{kr}-i\phi_{ki}\right)e^{-i\frac{2\pi}{\beta}k\tau}\right].
\end{equation}
Note that this partition function is strictly speaking not convergent
in the limit $k_{\text{co}}\rightarrow\infty$, due to the divergent
UV tail (which is why we explicitly keep the cutoff $k_{\text{co}}$
here). But the resulting heat capacity is well-defined for $k_{\text{co}}\rightarrow\infty$,
yielding
\begin{align*}
C_{v} & =\frac{1}{2}\sum_{k=-\infty}^{\infty}\left[\left(\frac{\frac{\beta}{c\eta}\left|k\right|+2\beta^{2}\frac{1}{lc}}{4\pi^{2}k^{2}+\frac{\beta}{c\eta}\left|k\right|+\frac{\beta^{2}}{lc}}\right)^{2}-\frac{2\frac{\beta^{2}}{lc}}{4\pi^{2}k^{2}+\frac{\beta}{c\eta}\left|k\right|+\frac{\beta^{2}}{lc}}\right].
\end{align*}
In the limit $1/l\rightarrow0$, this sum yields
\begin{equation}
C_{v}=\left(\frac{\beta}{4\pi^{2}c\eta}\right)^{2}\psi^{\left(1\right)}\left(\frac{\beta}{4\pi^{2}c\eta}\right),
\end{equation}
where $\psi^{\left(1\right)}$ is the first derivative of the Digamma
function. Assuming asymptotically large arguments, this function is
approximated as $\psi^{\left(1\right)}\left(z\right)\approx1/z+1/\left(2z^{2}\right)$,
yielding Eq.~\eqref{eq_heat_capacity} in the main text.

\section{Partial path integral method}\label{sec_partial_path_integral}
Here we derive the form of the partition function as shown in Eq. \eqref{eq_Z_PI} in the main text. We begin with the definition of a partition function
\begin{align}
    Z = \text{tr}\left[e^{-\beta H}\right],
\end{align}
where $H$ is the Hamiltonian of the full system (Eq. \eqref{eq_Lag_and_Hamil} in the main text). Since the full system is comprised of two subsystems: one with charge/phase degree of freedom and the other with the pseudo-spin degree of freedom, the total trace over the system can be written as the combination of traces over the two subsystems. Hence, we can write
\begin{align}
    Z = \text{tr}_{\sigma}\left[\text{tr}_{\phi}\left[e^{-\beta H}\right]\right],
\end{align}
where $\text{tr}_{\sigma}$ is the trace over the pseudo-spin degree of freedom and $\text{tr}_{\phi}$ is the trace over the charge/phase degree of freedom.

Our goal will be to write the trace $\text{tr}_{\phi}$ as a path integral, therefore we first replace this trace with the eigenstates $\left(\left|\phi\right>\right)$ of the phase operator, to get
\begin{align}
    Z = \int d\phi\text{tr}_{\sigma}\left[\left<\phi\right|e^{-\beta H}\left|\phi\right>\right].
\end{align}
Once we have the expression $\left<\phi\right|e^{-\beta H}\left|\phi\right>$, we can use the procedure similar to the one used to derive the path integral formulation of quantum mechanics \cite{Altland_Simons_book}. We begin by breaking the exponential of the Hamiltonian into $M$ components
\begin{align}
    \left<\phi\right|\left(e^{-\frac{\beta}{M} H}\right)^{M}\left|\phi\right> = \left<\phi\right|\underbrace{e^{-\Delta\tau H}\cdots e^{-\Delta\tau H}}_{M \text{ times}}\left|\phi\right>,
\end{align}
where we have defined $\Delta\tau=\beta/M$. Eventually we will take the limit $M\rightarrow\infty$, but for now we take it to be large enough so that we can use the following approximation
\begin{align}
    e^{-\Delta\tau H} = e^{-\Delta\tau \widehat{T}(N)}e^{-\Delta\tau V(\phi)}+\mathcal{O}(\Delta\tau^2).
\end{align}
Writing the exponential in the above factorized form allows us to diagonalize the kinetic energy term and the potential energy term independently. To use this independence we insert the resolution of identity ($\mathcal{I}$) in the expression $\left<\phi\right|e^{-\beta H}\left|\phi\right>$
\begin{align}
    \left<\phi\right|\left(e^{-\frac{\beta}{M} H}\right)^{M}\left|\phi\right> = \left<\phi\right|\mathcal{I}_{M}e^{-\Delta\tau \widehat{T}(N)}e^{-\Delta\tau V(\phi)}\mathcal{I}_{M-1}\cdots\mathcal{I}_{1}e^{-\Delta\tau \widehat{T}(N)}e^{-\Delta\tau V(\phi)}\left|\phi\right>,
\end{align}
where
\begin{align*}
    \mathcal{I}_{j} = \int d\phi_{j}\int dN_{j} \left|\phi_{j}\right>\left.\left<\phi_{j}\right|N_{j}\right>\left<N_{j}\right|\otimes\mathcal{I}_{\sigma}, 
\end{align*}
the subscript $j$ helps to differentiate the different phase and charge variables originating from different resolutions of identity, and $\mathcal{I}_{\sigma}$ is the identity in the pseudo-spin space.

Using the expression $\left.\left<\phi\right|N\right>=\left.\left<N\right|\phi\right>^{*}=e^{iN\phi}/\sqrt{2\pi}$, we can write
\begin{align}
    Z = \text{tr}_{\sigma}\left[\int \prod_{j=1}^{M}d\phi_{j}\int \prod_{j=1}^{M}\frac{dN_{j}}{2\pi} \mathcal{T}\exp{\left(-\Delta\tau\sum_{j=1}^{M}\left[\widehat{T}(N_{j})+V(\phi_{j-1})- iN_{j}\frac{\phi_{j}-\phi_{j-1}}{\delta\tau}\right]\right)}\right],
\end{align}
with the constraint that $\phi_{0}=\phi_{M}=\phi$. Here, we have also used the time ordering operator $\mathcal{T}$ because even if the charge and phase operators have now been replaced with their eigenvalues, the pseudo-spin degree of freedom is still quantum, and one has to arrange the quantum operators in the correst order of increasing imaginary time. We now use the specific form of $\widehat{T}(N)=(N+\lambda\sigma_{z})^{2}/2c + \gamma\sigma_{x}$, and perform the Gaussian integrals over the variables $N_{j}$ to get
\begin{align}
    Z = \text{tr}_{\sigma}\left[\int \prod_{j=1}^{M}\frac{cd\phi_{j}}{\delta\tau}\mathcal{T}\exp{\left(-\Delta\tau\sum_{j=1}^{M}\left[\frac{c}{2}\left(\frac{\phi_{j}-\phi_{j-1}}{\delta\tau}\right)^{2}+i\lambda\frac{\phi_{j}-\phi_{j-1}}{\delta\tau}\sigma_{z}+\gamma\sigma_{x}+V(\phi_{j-1}) \right]\right)}\right].
\end{align}
Finally we take the limit $M\rightarrow\infty$, while keeping $M\Delta\tau=\beta$ constant. This allows us to replace the sum $\sum_{j=1}^{M}\Delta\tau$ with the integral $\int_{0}^{\beta}d\tau$, and to also make the following replacements
\begin{align*}
    \frac{\phi_{j}-\phi_{j-1}}{\Delta\tau} &\rightarrow \partial_{\tau}\phi \equiv \dot{\phi}_{\tau}\\
    V(\phi_{j-1}) &\rightarrow V(\phi_\tau).
\end{align*}
Hence, for the concrete dipole model, the partition function becomes
\begin{align}
    Z = \text{tr}_{\sigma}\left[\oint D\phi\mathcal{T}\exp{\left(-\int_{0}^{\beta} d\tau\left[\frac{c}{2}\dot{\phi}_{\tau}^{2}+i\lambda\dot{\phi}_{\tau}\phi\sigma_{z}+\gamma\sigma_{x}+V(\phi) \right]\right)}\right],
\end{align}
where
\begin{align}
    \oint D\phi = \lim_{M\rightarrow\infty}\prod_{j=1}^{M}\int\frac{cd\phi_{j}}{\Delta\tau},
\end{align}
is the integral measure and the closed integral sign $\oint$ represents the fact that the integration is being performed over all the paths (in imaginary time) for which $\phi(\tau=0)=\phi(\tau=\beta)$. To bring this in the form of Eq. \eqref{eq_Z_PI}, we interchange the order of the path integral and the trace over the pseudo-spin degree of freedom. We also note that if we replace $t$ in the expression of $\widehat{T}^{*}(\dot{\phi})$ (Eq. \eqref{eq_Lag_and_Hamil} in the main text) by $-i\tau$, we get the final form of the partition function is
\begin{align}
    Z =\oint \mathcal{D}\phi \text{tr}_{\sigma}\left[\mathcal{T}e^{\int_0^\beta d\tau \left[\widehat{T}^*(i\dot{\phi}_\tau)-V(\phi_\tau)\right] }\right],
\end{align}
which, for the explicit dipole model, yields
\begin{align}
    T^{*}(i\partial_{\tau}\phi)= -\frac{c}{2}\dot{\phi}_{\tau}^{2}-i\lambda\dot{\phi}_{\tau}\sigma_{z}-\gamma\sigma_{x}.
\end{align}

\section{Noise power spectrum of bare LC resonator}\label{sec_noise_power}

In the main text, we provide the imaginary-time voltage noise power
spectrum for the bare LC resonator system either with [Eq.~\eqref{eq_Sv_diss}] or
without [Eq.~\eqref{eq_Sv_0}] an Ohmic resistor. We here derive these expressions.
We introduce the notation for averaging over paths of said LC resonator
as follows [in analogy to the ordinary path integral treatment, see
Eq.~\eqref{eq_supp_path_integral} above] 
\begin{equation}\label{eq_path_expectation_value}
\left\langle \bullet\right\rangle_{\text{path}}\equiv\lim_{k_{\text{co}}\rightarrow\infty}\frac{\prod_{k=0}^{k_{\text{co}}}\int d\phi_{k}e^{-\beta\left(\frac{c\omega_{k}^{2}}{2}+\frac{\omega_{k}}{4\pi\eta}+\frac{1}{2l}\right)\left|\phi_{k}\right|^{2}}\left(\bullet\right)}{\prod_{k=0}^{k_{\text{co}}}\int d\phi_{k}e^{-\beta\left(\frac{c\omega_{k}^{2}}{2}+\frac{\omega_{k}}{4\pi\eta}+\frac{1}{2l}\right)\left|\phi_{k}\right|^{2}}}.
\end{equation}
The above is a shorthand notation of the path integral introduced
in Eq.~\eqref{eq_supp_path_integral}, where for $k>0$, the integral $\int d\phi_{k}$
goes over the 2D complex plane, $\int d\phi_{kr}\int d\phi_{ki}$.
Plugging in Eq.~\eqref{eq_supp_phitau_Matsubara} for $\phi_\tau$, we can now calculate with ease the expectation value for
\begin{equation}
\left\langle \dot{\phi}_{\tau}\dot{\phi}_{\tau^{\prime}}\right\rangle _{\text{path}}=\frac{1}{2}\sum_{k=1}^{\infty}S_{k}\left[e^{i\omega_{k}\left(\tau-\tau^{\prime}\right)}+e^{-i\omega_{k}\left(\tau-\tau^{\prime}\right)}\right]
\end{equation}
with $S_{k}$ as given in Eq.~\eqref{eq_Sv_diss} of the main text. Equation~\eqref{eq_Sv_0}
follows of course from~\eqref{eq_Sv_diss} by setting $1/\eta=0$.

\section{Time dependent basis transformation and intrinsic UV cutoff}\label{sec_basis_transformation}

In the main text, we argue that the energy gap $\gamma$
provides an intrinsic UV cutoff. We here explain and demonstrate this
statement. The object we will be looking at is the partition function
$Z$, defined in Eq.~\eqref{eq_Z_PI} of the main text. In fact, we consider
it explicitly for the two-level toy model presented in the first part
of the main text, and take the ratio with respect to the bare LC resonator
partition function $Z_{0}$ (i.e., removing the additional quantum
dipole). This quantity can be expressed as
\begin{equation}
\frac{Z}{Z_{0}}=\text{tr}_{\sigma}\left[\left\langle \mathcal{T}e^{-\int_{0}^{\beta}d\tau\left(i\lambda\dot{\phi}_{\tau}\sigma_{z}+\gamma\sigma_{x}\right)}\right\rangle _{\text{path}}\right]
\end{equation}
where we use the notation for the average taken over the bare paths
as given above, in Eq.~\eqref{eq_path_expectation_value}. Now, there are two steps involved.

The first step consists of applying a time-dependent basis transformation
to the above path integral expression. We start from the time-ordered
propagator in Eq. (S27), and rewrite it explicitly in $K$ discrete
time slices of length $\Delta\tau=\beta/K$,
\begin{equation}
\mathcal{T}e^{-\int_{0}^{\beta}d\tau\left(i\lambda\dot{\phi}_{\tau}\sigma_{z}+\gamma\sigma_{x}\right)}=\lim_{K\rightarrow\infty}\prod_{j=1}^{K}e^{-\Delta\tau\left(i\lambda\dot{\phi}_{\tau_{j}}\sigma_{z}+\gamma\sigma_{x}\right)}=\prod_{j=1}^{K}e^{-\Delta\tau i\lambda\dot{\phi}_{\tau_{j}}\sigma_{z}}e^{-\Delta\tau\gamma\sigma_{x}},
\end{equation}
where it is important to keep the product $\prod$ ordered with ascending
$k$ index. The terms $e^{-\Delta\tau\gamma\sigma_{x}}$ can be recast
into the form of a time-dependent basis transformation
\begin{equation}
U_{j}=e^{-j\Delta\tau\gamma\sigma_{x}}
\end{equation}
such that
\begin{equation}
\prod_{j=1}^{K}e^{-\Delta\tau i\lambda\dot{\phi}_{\tau_{j}}\sigma_{z}}e^{-\Delta\tau\gamma\sigma_{x}}=U_{K}\prod_{j=1}^{K}e^{-\Delta\tau i\lambda\dot{\phi}_{\tau_{j}}U_{j}^{-1}\sigma_{z}U_{j}}.
\end{equation}
$U_{j}$ is not unitary, due to imaginary times. Going back to continuous
time, we get
\begin{equation}
\mathcal{T}e^{-\int_{0}^{\beta}d\tau\left(i\lambda\dot{\phi}_{\tau}\sigma_{z}+\gamma\sigma_{x}\right)}=e^{-\beta\gamma\sigma_{x}}\mathcal{T}e^{-i\lambda\int_{0}^{\beta}d\tau\dot{\phi}_{\tau}e^{\tau\gamma\sigma_{x}}\sigma_{z}e^{-\tau\gamma\sigma_{x}}}.
\end{equation}
At this stage, we already see that $\dot{\phi}_{\tau}$ is convoluted
with $e^{\pm\gamma\tau}$ terms, indicating already on a qualitative
level, that $\gamma$ likely acts as a cutoff energy scale for fluctuations
in $\dot{\phi}_{\tau}$.

In a second step, we make this intuitive insight quantitative, by
expanding in $\lambda$, and evaluating the correlators. For illustrative
purposes, we here focus on leading order (second order in $\lambda$),
where we get
\begin{equation}
\left\langle \text{tr}_{\sigma}\left[\mathcal{T}e^{-\int_{0}^{\beta}d\tau\left(i\lambda\dot{\phi}_{\tau}\sigma_{z}+\gamma\sigma_{x}\right)}\right]\right\rangle_\text{path} ^{\left(2\right)}=-\lambda^{2}e^{\beta\gamma}\int_{0}^{\beta}d\tau\int_{0}^{\tau}d\tau^{\prime}\left\langle \dot{\phi}_{\tau}\dot{\phi}_{\tau^{\prime}}\right\rangle_\text{path} \left[e^{2\left(\tau-\tau^{\prime}-\beta\right)\gamma}+e^{-2\left(\tau-\tau^{\prime}\right)\gamma}\right].
\end{equation}
Out of those two terms, the second one is the leading one, such that
we arrive at
\begin{equation}
\left\langle \text{tr}_{\sigma}\left[\mathcal{T}e^{-\int_{0}^{\beta}d\tau\left(i\lambda\dot{\phi}_{\tau}\sigma_{z}+\gamma\sigma_{x}\right)}\right]\right\rangle_\text{path} ^{\left(2\right)}\approx-\lambda^{2}e^{\beta\gamma}\sum_{k=1}^{\infty}\frac{\beta^{2}S_{k}}{2}\frac{\beta\gamma}{k^{2}\pi^{2}+\beta^{2}\gamma^{2}}.
\end{equation}
This explicitly shows that the sum over $k$ is cut off at energies
$\gg\gamma$.

\section{Derivation of large dipole-moment model}\label{sec_large_dipole_model}

In the main text, we discuss a second explicit model, which generalizes the two-level quantum dipole model to many dipoles. We here provide details on the derivation of the model, needed to reproduce Fig.~\ref{fig2}.

As pointed out in the main text, we generalize from one to many dipoles, where the total polarization operator is $\widehat{\Lambda}=\sum_k \lambda_k\sigma_z^{(k)}/2$. The intrinsic dynamics $\widehat{\Gamma}$ still includes a tunneling between charge (dipole) configurations, but includes now in addition a dipole-dipole interaction $\Gamma=\sum_k \gamma_k\sigma_x^{(k)}/2+\sum_{k,k^\prime}J_{k,k^\prime}\sigma_z^{(k)}\sigma_z^{(k^\prime)}/4$. This model can be easily solved under the simplifying assumption that all dipole moments be identical and oriented the same way with respect to the capacitor ($\gamma_k=\gamma$, $\lambda_k=\lambda$) and, in addition, let the dipole-dipole interaction not depend on the spatial separation between dipoles $J_{k,k^\prime}=J$. Thus, Eq.~\eqref{eq_general_model} in the main text depends only on the total pseudo-spins, $S_\alpha=\sum_k\sigma_\alpha^{(k)}/2$ ($\alpha=x,y,z$), where the pseudo-spin magnitude (total dipole magnitude) $S^2=S_x^2+S_y^2+S_z^2$ is a conserved quantity.  Fixing the total pseudo-spin to the maximal value $S=K/2$, we can represent the large spin in the standard representation
\begin{align}
    S_z&=\sum_{m=-S}^S m\vert m\rangle \langle m\vert\ , \\
    S_x&=\frac{1}{2}\sum_{m=-S}^{S-1} \sqrt{(S+m+1)(S-m)}\vert m+1 \rangle \langle m\vert+\text{h.c.}
\end{align}
This matrix representation is used in generating the plots of Fig.~\ref{fig2}.